\documentclass[twocolumn,showpacs,preprintnumbers,amsmath,amssymb]{revtex4}
\usepackage{graphicx,xcolor}
\usepackage{dcolumn}% Align table columns on decimal point
\usepackage{bm}% bold math

\begin{document}

\preprint{APS/123-QED}

\title{Low-lying level structure of $\Lambda$ hypernuclei and spin dependence of $\Lambda N$ interaction with antisymmetrized molecular dynamics}% Force line breaks with \\

\author{Masahiro Isaka$^1$, Yasuo Yamamoto$^2$, and Toshio Motoba$^{3,4}$}
\affiliation{$^1$ Science Research Center, Hosei University, 2-17-1 Fujimi, Chiyoda, Tokyo 102-8160, Japan}
\affiliation{$^2$ RIKEN Nishina Center, Wako, Saitama 351-0198, Japan}
\affiliation{$^3$ Laboratory of Physics, Osaka Electro-Communication University, Neyagawa 572-8530, Japan}
\affiliation{$^4$ Yukawa Institute for Theoretical Physics, Kyoto University, Kyoto 606-8502, Japan}

%\author{Ann  Author}
%\affiliation{%
%\affiliation{

\date{\today}% It is always \today, today,

\begin{abstract}
$\Lambda N$ spin-spin and spin-orbit splittings in low-lying excitation spectra
are investigated for $p$-shell $\Lambda$ hypernuclei
on the basis of the microscopic structure calculation
within the antisymmetrized molecular dynamics,
where the $\Lambda N$ $G$-matrix interaction
derived from the baryon-baryon interaction model ESC (extended soft core) is used.
It is found that the ground-state spin-parity is systematically reproduced
in the $p$-shell $\Lambda$ hypernuclei
by tuning the $\Lambda N$ spin-spin and spin-orbit interactions
so as to reproduce the experimental data of $^{4}_\Lambda$H, $^{7}_\Lambda$Li and $^9_\Lambda$Be.
Furthermore, we also focus on the excitation energies of the excited doublets
as well as the energy shifts of them by the addition of a $\Lambda$ particle.
\end{abstract}

\pacs{Valid PACS appear here}% PACS, the Physics and Astronomy
\maketitle
\section{Introduction}

Hypernuclear physics makes substantial progress by both theoretical and experimental works in the last decades.
In particular, experimental information of $\Lambda$ hypernuclei has been increased by the counter experiments such as $(\pi^+, K^+)$ reactions \cite{PPNP57.564(2006)}.
Combined with the $\gamma$-ray spectroscopy techniques, the low-lying level structure of $\Lambda$ hypernuclei have been revealed precisely.
Recently, the $(e, e'K^+)$ reaction experiments have been developed at Thomas Jefferson Laboratory (JLab),
where the absolute values of binding energies are expected to be measured in a wide mass region including medium-heavy hypernuclei with high resolution.
These data of hypernuclei are essential to understand hyperon-nucleon ($YN$) interactions,
because $YN$ scattering experiments are rather difficult due to short life-time of hyperons.
For example, from the analysis of the binding energies of $\Lambda$ hypernuclei, $B_\Lambda$,
one can evaluate the depth of the single-particle potential of a $\Lambda$ particle, $U_\Lambda$,
in nuclear matter.
In associated with the developments on baryon-baryon interaction models
\cite{PRC40.2226(1989),PRC59.21(1999),PRC73.044008(2006),PTPS185.14(2010),NPA500.485(1989),NPA570.543(1994),PPNP58.439(2007)},
an important role of the $\Lambda N - \Sigma N$ coupling to give a reasonable $U_\Lambda$ value
in nuclear matter has been revealed.

Based on the precise data of $\Lambda$ hypernuclei, structure of hypernuclei is also investigated and becomes one of the important issues.
Especially, structure of light $\Lambda$ hypernuclei is of interest
because ordinary nuclei in $p$-shell and $sd$-shell mass regions have various structures such as cluster and deformed mean-field like structures near the ground states.
For example, in $^7_\Lambda$Li \cite{PTP70.189(1983),PTPS81.42(1985),PRC59.2351(1999),PRL86.1982(2001)},
it was predicted that the addition of a $\Lambda$ particle reduces the intercluster distance between $\alpha$ and $d$ of the core nucleus $^6$Li,
which was confirmed through the observations of the $B(E2)$ values by the $\gamma$-ray spectroscopy experiment.
In the other $\Lambda$ hypernuclei such as $^{13}_\Lambda$C,
several theoretical calculations showed that a $\Lambda$ particle reduces the nuclear quadrupole deformation of the ground states
\cite{PRC76.034312(2007),PRC78.054311(2008),PTP123.569(2010),PRC83.044323(2011),PRC84.014328(2011),PRC90.064302(2014),PRC91.024327(2015),PRC91.054306(2015)}.
In addition, the coexistence of pronounced cluster and mean-field like structures in core nuclei can cause changes of level ordering in excitation spectra,
which were predicted in $p$- and $sd$-shell $\Lambda$ hypernuclei such as $^{13}_\Lambda$C \cite{PRL85.270(2000)} and $^{21}_\Lambda$Ne \cite{PRC83.054304(2011)}.
Furthermore, it was pointed out that size differences between the ground and excited states in $p$-shell nuclei
lead to energy shifts of excited states in the corresponding hypernuclei \cite{PRC97.024330(2018)}.

The structure of core nuclei also affects the systematics of $B_\Lambda$ values
through the density dependence of the $\Lambda$-nucleon ($\Lambda N$) interaction,
which was investigated by the authors (M. I. and Y.Y) with structure calculations
within the framework of the antisymmetrized molecular dynamics for hypernuclei (HyperAMD)
adopting the $G$-matrix interactions (called YNG) as a $\Lambda N$ effective interaction \cite{PRC94.044310(2016),PRC95.044308(2017)}.
The YNG interactions are obtained by the $G$-matrix calculation in nuclear matter from the Nijmegen
Extended Soft-Core (ESC) potentials, which depend on the nuclear Fermi momentum $k_F$.
In Refs. \cite{PRC94.044310(2016),PRC95.044308(2017)}, the authors succeeded in reproducing the existing $B_\Lambda$ data as a function of mass number $A$,
namely the mass dependence of $B_\Lambda$, where the averaged density approximation (ADA) is applied to treat the $k_F$ dependence.
This is achieved by the developments of the interaction models and microscopic calculation of hypernuclear structure.
In Ref. \cite{PRC95.044308(2017)}, it was also found that fine tuning of the $\Lambda N$ YNG interactions is still necessary
to describe the level ordering of the ground-state spin doublet partners properly,
which is one of the purposes of the present paper.

The level ordering of the spin doublets is mainly determined by properties of spin-dependent parts
of $\Lambda N$ interactions, namely the $\Lambda N$ spin-spin and spin-orbit interactions,
because the spin doublet states in $\Lambda$ hypernuclei are generated by the coupling between the non-zero spin of the core states
and the $\Lambda$ particle with spin $1/2$.
For the spin-spin part, the historically important data were the spin doublet ($0^+,1^+$) states of
$^4_\Lambda$He ($^4_\Lambda$H). These hypernuclear data were first used for the parameter fitting
of the hyperon-nucleon one-boson exchange model NSC97 \cite{PRC59.21(1999)}.
Among 6 versions (a,b,c,d,e,f) with different spin-spin interactions, NSC97f was found to
reproduce reasonably the above data of spin doublet splitting.
Furthermore, the spin doublet states of $^{12}_{\Lambda}$C and
$^{11}_{\Lambda}$B were shown to be described well by the shell-model
analysis with the $G$-matrix interaction derived from NSC97f \cite{Yam2010}.
On the other hand, spin-orbit splittings also were long-standing problems in hypernuclear
physics. Qualitatively, small values of splitting energies are obtained by large cancellations
between symmetric LS (SLS) and anti-symmetric LS (ALS) contributions. In the case of using the
$G$-matrix interaction from NSC97f, however, the cancellation was shown to be not enough to
reproduce quantitatively the small value of the experimental data \cite{NPA804.99(2008),Hiyama10}.

The ESC model~\cite{PTPS185.14(2010)} was proposed in order to improve important deficiencies
of NSC97, where two-meson and meson-pair exchanges were taken into account explicitly instead
of ``effective bosons" in one-boson-exchange models such as NSC97. In the ESC models, then,
the spin-spin parts in even states were designed to give similarly those of NSC97f,
and the SLS and ALS parts were done to cancell more substantially than those of NSC97f.
The shell-model analysis for spin doublet splittings in $^{12}_{\Lambda}$C and $^{11}_{\Lambda}$B
were performed by using the $G$-matrix interactions derived from the ESC models \cite{Yam2010}.
The obtained values of splitting energies were rather larger than those for NSC97f, namely than
the experimental values, because the odd-state interactions in the former were rather different
from those in the latter. These results are consistent with the incorrect ordering of the
doublet partners in Refs. \cite{PRC95.044308(2017)}.

The aim of this paper is to study spin-spin and spin-orbit splittings in excitation spectra of $\Lambda$ hypernuclei systematically.
For the present purpose we will especially investigate necessary corrections
in both even- and odd-state spin-dependent parts of the YNG interactions
so as to reproduce the excitation spectra of $p$-shell $\Lambda$ hypernuclei.
Thus, we will finally propose a revised version of the YNG interactions
which can be applicable to whole mass regions of hypernuclei.

This paper is organized as follows.
In the next Section, the theoretical framework of HyperAMD is explained.
In Sec. III, we discuss the spin-doublet splittings obtained in a schematic
model in order to suggest leading properties in spherical hypernuclei.
In Sec. IV, we discuss the effects of the $\Lambda N$ spin-spin and spin-orbit interactions
by showing the excitation spectra of the hypernuclei as well as those of the core nuclei.
Section V summarizes this work.

\section{Theoretical framework}

In this paper, we apply an extended version of the antisymmetrized molecular dynamics for hypernuclei named HyperAMD
to several $p$-shell $\Lambda$ hypernuclei basically following Refs. \cite{PRC94.044310(2016),PRC95.044308(2017)}.
This model enables us to describe various nuclear structures and to investigate the dynamical changes 
caused by the addition of a $\Lambda$ particle in the hypernuclei around the $p$-$sd$ shell regions.

\subsection{Hamiltonian}

The Hamiltonian used in this study is given as
\begin{eqnarray}
H = T_{N}+ T_{\Lambda}- T_{g}+ V_{NN}+ V_{C}+ V_{\Lambda N},
\end{eqnarray}
where $T_{N}$, $T_{\Lambda}$, and $T_{g}$ are the kinetic energies of the nucleons, $\Lambda$ particle, and center-of-mass motion, respectively.
$V_{NN}$ is the effective nuclear force.
The Coulomb interaction $V_{C}$ is approximated by the sum of seven Gaussians.

We use the Gogny D1S force \cite{Gogny1,Gogny2} as the effective nuclear force $V_{NN}$.
One of the characteristics of Gogny D1S is the density-dependent interaction acting as a repulsive force at high density, which is essential to describe nuclear saturation property.
As discussed in Refs. \cite{PRC94.044310(2016),PRC95.044308(2017)}, for reproducing the $B_\Lambda$ values systematically,
it is indispensable to describe structure of core nuclei, especially core deformation, properly.
The Gogny D1S force is one of the effective interactions
that gives better description of nuclear deformation and reasonable values of the nuclear binding energies
for the ground states of normal nuclei in a wide mass region.
In this study, since we focus on the effects by the spin-dependent part of the $\Lambda N$ interaction,
it is necessary to describe excitation spectra of the core nuclei properly as well as possible.
Though the Gogny D1S describes structure of the ground states,
it may give small deviation of energies for excited states from the observed data.
In such cases, for the quantitative discussion, we slightly change the parameter sets of the Gogny D1S
or use the Volkov No. 2 force \cite{Volkov2} instead, which is discussed in Sec. \ref{IIIA} in detail.

\subsection{$\Lambda N$ interaction}

As for the $\Lambda N$ effective interaction $V_{\Lambda N}$,
we use the the $G$-matrix interaction derived from the Nijmegen ESC potential, ESC14,
where the $\Lambda N$-$\Sigma N$ coupling is renormalized by the $G$-matrix calculation.
As shown in Ref. \cite{PRC95.044308(2017)}, using the ESC14 force with the many-body effects (ESC14+MBE),
the HyperAMD calculation nicely reproduces the existing data of $B_\Lambda$ systematically.
However, level ordering of spin doublet partners is incorrect in several $p$-shell $\Lambda$ hypernuclei.
In this paper, we investigate spin-spin and spin-orbit splittings of the $p$-shell $\Lambda$ hypernuclei using ESC14+MBE,
where the $k_F$ dependence of the $G$-matrix interaction is treated by the ADA procedure in the same way as in Refs. \cite{PRC94.044310(2016),PRC95.044308(2017)}.
Furthermore, we also apply the NSC97f for the comparison.
%%%%%% added by yy
In the case of NSC97f, its strong odd-state repulsions play an important role to reproduce
the mass dependence of $B_\Lambda$ instead of the above MBE.

The $\Lambda N$ $G$-matrix interaction $ V_{\Lambda N}$ is composed of the central ($V^{\rm cent}_{\Lambda N}$) and spin-orbit ($V^{\rm LS}_{\Lambda N}$) forces,
namely $V_{\Lambda N} = V^{\rm cent}_{\Lambda N} + V^{\rm LS}_{\Lambda N}$.
The $\Lambda N$ central force $V^{\rm cent}_{\Lambda N}$ is written as,
\begin{eqnarray}
\nonumber
V^{\rm cent}_{\Lambda N} &=& v^{(^1 E)} \hat{P}(^1 E) + v^{(^3 E)} \hat{P}(^3 E) \\
               &\ & + v^{(^1 O)} \hat{P}(^1 O) + v^{(^3 O)} \hat{P}(^3 O),
\label{eq:LN}
\end{eqnarray}
where
\begin{eqnarray}
\label{eq:YNGe}
v^{(c)} (k_F, r) &=& \sum^{3}_{i=1} ( a^{(c)}_i + b^{(c)}_i k _F + c^{(c)}_i k_F^2 ) e^{ -r^2 /\beta^2_i},\\
\label{eq:YNGc}
c &=& \ ^1 E,\ ^3 E,\ ^1 O\ {\rm or}\ ^3 O.
\end{eqnarray}
The parameters $a^{(c)}_i$, $c^{(c)}_i$ and $c^{(c)}_i$ of ESC14+MBE are shown in Tab. \ref{Tab:central} for all channels,
which are the same as used in Ref. \cite{PRC95.044308(2017)}.
Using the Pauli's spin matrix $\vec{\sigma}$, one can rewrite $ V^{\rm cent}_{\Lambda N}$ as
\begin{eqnarray}
\nonumber
V^{\rm cent}_{\Lambda N} &=& \sum_{i=1}^3 \{
  ( v^{E(i)}_0 + v^{E(i)}_\sigma \vec{\sigma} \cdot \vec{\sigma}) \hat{P}(E) \\
 &\ & +  ( v^{O(i)}_0 + v^{O(i)}_\sigma \vec{\sigma} \cdot \vec{\sigma}) \hat{P}(O)
\} e^{-r^2/\beta_i^2},
\label{eq:VevenOdd}
\end{eqnarray}
where $\hat{P}(E)$ and $\hat{P}(O)$ are the projectors for the even and odd parity states, respectively.
Thus, the spin-spin even and odd parity interactions act through the $\Lambda N$ central force.

The $\Lambda N$ spin-orbit force $ V^{\rm LS}_{\Lambda N}$ consists of the symmetric
and asymmetric LS (SLS and ALS) components.
Each of SLS and ALS of ESC14+MBE is also given by the same functional form as Eq.(\ref{eq:YNGe}) with
$c = SLS$ or $ALS$, where the parameters $a^{(c)}_i$, $b^{(c)}_i$, $c^{(c)}_i$ and $\beta_i$ are listed in Tab. \ref{Tab:LS}.

In our $G$-matrix interaction, the $\Lambda N$-$\Sigma N$ coupling interactions of ESC14 
are renormalized into the effective central interactions. 
It is possible, then, to obtain residual $\Lambda N$-$\Lambda N$ and $\Lambda N$-$\Sigma N$ interactions 
composed of central and tensor parts, though they are not used in this work. 
In Ref.\cite{Yam2010} such residual interactions were obtained for some versions of the ESC model, 
to which those for ESC14 are more or less similar.

\subsection{Wave Function}

The variational wave function of a single $\Lambda$ hypernucleus is described by the parity-projected wave function,
$\Psi^{\pm }= P^{\pm} A \{ \varphi_1, \cdots, \varphi_A \} \otimes \varphi_\Lambda$, where
\begin{eqnarray}
\varphi _{i} \propto e^{ - \sum_{\sigma} \nu_{\sigma} ( r_{\sigma} - \vec{Z}_{i \sigma} )^2 }
 \otimes (u_{i} \chi_{\uparrow}+ v_{i} \chi_{\downarrow })\otimes ( {\rm p} \ {\rm or} \ {\rm n} ), \\
\varphi_{\Lambda} \propto \sum_{m=1}^{M} c_{m} e^{-\Sigma_{\sigma} \nu_{\sigma} ( r_{\sigma} - \vec{z}_{m \sigma} )^2 }
 \otimes ( a_{m} \chi_{\uparrow}+ b_{m} \chi_{\downarrow} ).
\end{eqnarray}
The single-particle wave packet of a nucleon $\varphi_{i}$ is described by a single Gaussian,
whereas that of $\Lambda$, $\varphi_{\Lambda}$, is represented by a superposition of Gaussian wave packets.
The variational parameters $\vec{Z}_{i}$, $\vec{z}_{m}$, $\nu_\sigma$, $u_{i}$, $v_{i}$, $a_{m}$, $b_{m}$,
and $c_{m}$ are determined to minimize the total energy under the constraint
on the nuclear quadrupole deformation $(\beta, \gamma)$,
and the optimized wave function $\Psi^\pm (\beta, \gamma)$ is obtained for each given $(\beta, \gamma)$.

After the energy variation, we project out the eigenstate of the total angular momentum $J$ for each set of $(\beta, \gamma)$,
\begin{eqnarray}
\Psi^{J \pm}_{MK} (\beta, \gamma) = \frac{2J+1}{(8 \pi)^2}
\int d \Omega D^{J \ast}_{MK} (\Omega) R ( \Omega ) \Psi^{\pm} (\beta, \gamma).
\end{eqnarray}
The numerical integration is performed for the three Euler angles $\Omega$.
To obtain both the ground and excited states of hypernuclei, we also perform the generator coordinate method (GCM) calculation,
namely superposition of the different $K$ and $(\beta, \gamma)$ values as
\begin{eqnarray}
\Psi_{n}^{J \pm} = \sum_p \sum_{ K = - J }^J c_{npK} \Psi^{J \pm}_{MK} (\beta_p, \gamma_p),
\end{eqnarray}
where $n$ represents quantum numbers other than total angular momentum and parity.
The coefficients $c_{npK}$ are determined by solving the Griffin-Hill-Wheeler equation.

%%%%%%%%%%%%%%%%%%%%%%%%

\begin{table}
  \caption{
   Parameters of the $\Lambda N$ central force of ESC14+MBE represented by the three range Gaussian forms given in Eq. (\ref{eq:YNGe}).
   $a^{(c)}_i$ [MeV], $b^{(c)}_i$ [MeV $\cdot$ fm], $c^{(c)}_i$ [MeV $\cdot$ fm$^2$] and $\beta$ [fm] are given for each $i$.
   Parameters after the tuning done in Sec. \ref{IIIB} are also shown in parenthesis.
   }
  \label{Tab:central}
  \begin{ruledtabular}
  \begin{tabular}{ccccc}
         &         $i$ &       1 &                2 &     3 \\
  $c$    &   $\beta_i$ &    0.50 &             0.90 &  2.00 \\
  \hline
  $^1E$  & $a^{(^1E)}_i$ & -3434.0 &   416.71(413.11) & -1.708 \\
         & $b^{(^1E)}_i$ &  6937.0 &         -1108.74 &  0.0 \\
         & $c^{(^1E)}_i$ & -2635.0 &           444.74 &  0.0 \\
  \\
  $^3E$  & $a^{(^3E)}_i$ & -1933.0 &   214.56(215.76) & -1.295 \\
         & $b^{(^3E)}_i$ &  4698.0 &          -782.11 &  0.0 \\
         & $c^{(^3E)}_i$ & -1974.0 &            357.4 &  0.0 \\
  \\
  $^1O$  & $a^{(^1O)}_i$ &   206.1 &        94.2(4.2) & -0.8292 \\
         & $b^{(^1O)}_i$ &  -30.52 &           -39.47 &  0.0 \\
         & $c^{(^1O)}_i$ &   16.23 &           66.481 &  0.0 \\
  \\
  $^3O$  & $a^{(^3O)}_i$ &  2327.0 & -229.15(-199.15) & -0.9959 \\
         & $b^{(^3O)}_i$ & -2361.0 &           130.68 &  0.0 \\
         & $c^{(^3O)}_i$ &   854.3 &            23.02 &  0.0 \\
  \end{tabular}
  \end{ruledtabular}
\end{table}

\begin{table}
  \caption{
  Same as Tab. \ref{Tab:central} but for SLS and ALS represented by the three range Gaussian forms given in Eq. (\ref{eq:YNGe}).
  }
  \label{Tab:LS}
  \begin{ruledtabular}
  \begin{tabular}{ccccc}
       &  $i$ & 1 & 2 & 3 \\
  $c$  &  $\beta_i$ & 0.40 & 0.80 & 1.20 \\
  \hline
  $SLS$  & $a^{\rm (SLS)}_i$ & -12920.0 &  372.4 & -2.030 \\
         & $b^{\rm (SLS)}_i$ &  24580.0 & -840.0 &  0.0 \\
         & $c^{\rm (SLS)}_i$ & -10180.0 &  337.1 &  0.0 \\
  \hline
  $ALS$  & $a^{\rm (ALS)}_i$ &  1985.0 &  12.73 & 2.109 \\
         & $b^{\rm (ALS)}_i$ & -1828.0 &  41.30 & 0.0 \\
         & $c^{\rm (ALS)}_i$ &   679.8 & -17.58 & 0.0 \\
  \end{tabular}
  \end{ruledtabular}
\end{table}

\section{Doublet Energy Splitting in a Schematic Model}
%%%%%%%%%%%%%%%%%%%%%%%%%%%%%%%

Before going to discuss results of the detailed HyperAMD
calculations, we point out some schematic relations among
the ground-state spin-doublet splittings in hypernuclei that
are obtained on the shell model basis, rather independently
of the other sections. Then, secondly, we will show
the share of odd-state interaction components in these
doublets also in a schematic way without entering into
detailed forms of the $\Lambda N$ interactions.

Noting that the typical hypernuclei
such as $^{10}_{\,\Lambda}\mbox{Be}$,
$^{10}_{\,\Lambda}\mbox{B}$, $^{11}_{\,\Lambda}\mbox{B}$,
$^{12}_{\,\Lambda}\mbox{B}$, and $^{12}_{\,\Lambda}\mbox{C}$
are dominantly of spherical nature, we take a schematic
prescription specifically in this section.
Here we take the $L$-$S$ coupling shell model scheme  and
assume that each ground state  of the corresponding nuclear
core $(J_c,T_c)$ is described by an SU(3)
configuration~\cite{Elliot1958, Harvey1968} with
maximum orbital symmetry, denoted by $[f](\lambda,\mu)$.
Then hypernuclear spin-doublet wave functions
$\Psi_A(J=J_c\pm \tfrac{1}{2})$ for
$^{10}_{\,\Lambda}\mbox{Be}(J=1^-,2^-)$,
$^{11}_{\,\Lambda}\mbox{B}(5/2^+,7/2^+)$,
and   $^{12}_{\,\Lambda}\mbox{B}(1^-,2^-)$, are described
respectively as
\begin{eqnarray}
 & \bigl\{ ^9\mbox{Be};
  p^5[41](31)_{L=1}(S=\tfrac{1}{2}, T_c=\tfrac{1}{2}),
   J_c=3/2^-_g \bigr\} \times s_{1/2}^{\Lambda} \,, \qquad \nonumber\\
 & \bigl\{^{10}\mbox{B}; p^6[42](22)_{L=2}(S=1, T_c=0),
   J_c=3^+_g \bigr\} \times s_{1/2}^{\Lambda} \,,   \qquad \nonumber\\
 & \bigl\{^{11}\mbox{B};
  p^7[41](31)_{L=1}(S=\tfrac{1}{2}, T_c=\tfrac{1}{2}),
   J_c=3/2^-_g \bigr\} \times s_{1/2}^{\Lambda} \,. \qquad \nonumber
\end{eqnarray}
The analogous wave functions are given for
 $^{10}_{\,\Lambda}\mbox{B}(1^-,2^-)$ and
 $^{12}_{\,\Lambda}\mbox{C}(1^-,2^-)$, respectively.
 Hereafter $T_c$ is implicit, as the $\Lambda$ isospin is zero.

In order to estimate the hypernuclear doublet energy splitting,
the total $\Lambda N$ interaction energy defined by
\begin{equation}
 E_A(J=J_c\pm \tfrac{1}{2}) \equiv
  \big\langle \Psi_A(J) \big| \sum_i V_{N\Lambda}(i,1)  \big|\,
  \Psi_A(J) \big\rangle
\end{equation}
is calculated analytically within the full $p$-shell $L$-$S$ coupling
model space.  The resultant total energy is expressed in
terms of the $\Lambda N$ two-body interaction matrix elements
in the $jj$-coupling instead of the $L$-$S$ coupled one
$\langle p^N s^{\Lambda}(L)S | v | p^N s^{\Lambda}(L')S'\rangle_{J'}$,
because here we intend to show the share of $p_{3/2}$
and $p_{1/2}$ contributions. For this purpose,  we use the following
simplified notations.
\begin{eqnarray}
  &\ V_3(J') \equiv \langle p^N_{3/2} s_{1/2}^{\Lambda} | v
       | p^N_{3/2} s_{1/2}^{\Lambda}\rangle_{J'=1^-,2^-}, \nonumber \\
  &\ V_1(J') \equiv \langle p^N_{1/2} s_{1/2}^{\Lambda} | v
        | p^N_{1/2} s_{1/2}^{\Lambda}\rangle_{J'=0^-,1^-},    \\
  & V_{31}(J') \equiv  \langle p^N_{3/2} s_{1/2}^{\Lambda} | v
       | p^N_{1/2} s_{1/2}^{\Lambda}\rangle_{J'=1^-}. \nonumber
\end{eqnarray}
For the interaction energies $E_{10}(J)$ of
$^{10}_{\,\Lambda}\mbox{Be}(1^-,2^-)$, the detailed
calculation leads to the following expressions:
\begin{eqnarray}
E_{10}(1^-) &=& \tfrac{\,48\,}{45} V_3(2^-) + \tfrac{\,5\,}{9} V_3(1^-)
  +\tfrac{\,5\,}{9} V_1(1^-)      \nonumber \\
  &+& \tfrac{ 1 }{\,3\,}V_1(0^-)  - \tfrac{9\sqrt{2}}{5} V_{31}(1^-),         \\
E_{10}(2^-) &=& \tfrac{\,501\,}{225} V_3(2^-) + \tfrac{\,16\,}{25} V_3(1^-)
  +\tfrac{\,37\,}{25} V_1(1^-)   \nonumber \\
  &+& \tfrac{\,147\,}{225}V_1(0^-)  + \tfrac{27\sqrt{2}}{25} V_{31}(1^-).
\end{eqnarray}
Then we get the doublet energy splitting $\varDelta E_{10}$ as
\begin{eqnarray}
  \varDelta E_{10}(2^- -1^-_g) =
   \frac{29}{\,25\,}\{V_3(2^-) - V_3(1^-)\}  \qquad \nonumber \\
   \qquad +\ \frac{8}{\,25\,} \{V_1(0^-) - V_1(1^-)\}
    +\frac{72\sqrt{2}}{25} V_{31}(1^-). 
\label{Eq:A10split}
\end{eqnarray}
The energies of $^{12}_{\,\Lambda}\mbox{B}(1^-,2^-)$ are
obtained as follows:
\begin{eqnarray}
E_{12}(1^-) &=& \tfrac{\,149\,}{60} V_3(2^-) + \tfrac{\,53\,}{20} V_3(1^-)
  +\tfrac{\,8\,}{5} V_1(1^-)       \nonumber \\
  &+& \tfrac{\ 4\ }{15}V_1(0^-)  - \tfrac{9\sqrt{2}}{5} V_{31}(1^-),         \\
E_{12}(2^-) &=& \tfrac{1093}{300} V_3(2^-) + \tfrac{\,149\,}{100} V_3(1^-)
  +\tfrac{\,32\,}{25} V_1(1^-)   \nonumber \\
  &+& \tfrac{\,44\,}{75}V_1(0^-)  + \tfrac{27\sqrt{2}}{25} V_{31}(1^-).
\end{eqnarray}
As a result of the difference $E_{12}(2^-)-E_{12}(1^-)$, it is quite interesting to obtain exactly the same energy
splitting of $\varDelta E_{12}$ for $^{12}_{\Lambda}\mbox{B}$  as that for $^{10}_{\,\Lambda}\mbox{Be}$
given by Eq.(\ref{Eq:A10split}). Thus we can write 
\begin{eqnarray}
 \varDelta E_{12}(2^- -1^-_g) =  \varDelta E_{10}(2^- -1^-_g).
\label{Eq:A12split}
\end{eqnarray}
The doublet energies for
$^{11}_{\,\Lambda}\mbox{B}(5/2^+,7/2^+)$ are obtained in the similar
way as
\begin{eqnarray}
E_{11}(5/2^+) &=& \tfrac{\,25\,}{18} V_3(2^-) + \tfrac{\,8\,}{ 3 } V_3(1^-)
  +\tfrac{\,31\,}{18} V_1(1^-)       \nonumber \\
  &+& \tfrac{\ 5\ }{18}V_1(0^-)  - \tfrac{16\sqrt{2}}{9} V_{31}(1^-),         \\
E_{10}(7/2^+) &=& \tfrac{10}{\ 3\ } V_3(2^-) + \tfrac{\ 3\ }{ 3 } V_3(1^-)
  +\tfrac{\ 4\ }{ 3 } V_1(1^-)   \nonumber \\
  &+& \tfrac{\ 2\ }{ 3 }V_1(0^-)  + \tfrac{12\sqrt{2}}{ 9 } V_{31}(1^-).
\end{eqnarray}
Therefore the doublet energy splitting  $\varDelta E_{11}$ is given by
\begin{eqnarray}
 \varDelta E_{11}(7/2^+ - 5/2^+_g) =
   \frac{35}{\,18\,}\{V_3(2^-) - V_3(1^-)\}  \quad  \nonumber \\
   \quad +\ \frac{ 7 }{\,18\,} \{V_1(0^-) - V_1(1^-)\}
    +\frac{28\sqrt{2}}{ 9 } V_{31}(1^-).
\label{Eq:A11split}
\end{eqnarray}
We have two experimental energy splittings
so far \cite{Miura2005,Hosomi2015} among the adjacent
hypernuclei concerned here:
\begin{eqnarray}
 \varDelta E(^{12}_{\,\Lambda}{\rm C}\,;2^- -1^-_g)^{\rm exp}
      =0.162 \ {\rm MeV} ,    \\
 \varDelta E(^{11}_{\,\Lambda}{\rm B};7/2^+ -5/2^+_g)^{\rm exp}
    =0.263 \ {\rm MeV} .
\label{Eq:ExpC12B11split}
\end{eqnarray}
If we compare the leading terms of  Eq.(\ref{Eq:A11split})  and
Eq.(\ref{Eq:A12split}), we get the ratio
 $\varDelta E_{11}/\varDelta E_{12} \simeq
 (\tfrac{35}{18})/(\tfrac{29}{25})\simeq 1.67$, since the remaining
 terms amount to much smaller contributions.
 This theoretical ratio is in very good agreement with
 the experimental ratio ( 1.62 ). This fact manifests certain
 applicability of the present prescription based on the SU(3)
 classification.  Thus, on the basis of the relation obtained by
 Eq.(\ref{Eq:A12split}), we firmly predict the unknown
 energy splittings between the spin-doublet states as
\begin{equation}
\varDelta E(^{10}_{\,\Lambda}\mbox{Be}) \simeq
\varDelta E(^{10}_{\,\Lambda}\mbox{B}) \simeq
\varDelta E(^{12}_{\,\Lambda}\mbox{B}) \simeq
\varDelta E(^{12}_{\,\Lambda}\mbox{C})^{\rm exp}.
\end{equation}
In the recent $^{12}\mbox{C}(e,e'K^+)^{12}_{\,\Lambda}\mbox{B}$
experiment, a spectroscopic analysis of the lowest peak shape
has been made, suggesting
 $\varDelta E(^{12}_{\,\Lambda}\mbox{B}) \simeq 0.18$ MeV
 \cite{Tang2014}, which is well compared with 
 $\varDelta E(^{12}_{\,\Lambda}\mbox{C})^{\rm exp}$ 
 and is also consistent with a theoretical  prediction of 
 the energies and cross sections \cite{Motoba2017}.

In the actual calculations with the HyperAMD framework, as
shown in the next section, the results depend on the detailed
realistic structures such as hypernuclear deformation and
density distribution. Nevertheless the basic relations
mentioned above give a useful insight into these hypernuclear
structures.

Second, it is also interesting to see the contribution from relative
odd-state interactions to the doublet splitting. By expanding the
two-body interaction matrix elements
in terms of the relative and CM coordinates, the splitting energy
for the case of $\varDelta E_{10}$ is reexpressed as
\begin{eqnarray}
&&\hspace{-0.5cm}\varDelta E_{10}(2^- -1^-_g) \nonumber \\
 &=& \frac{7}{\,5\,} \langle 0\tilde{s} |V^{(^3E)}(r)| 0\tilde{s} \rangle
  +\frac{13}{\,15\,} \langle 0\tilde{s} |V^{(^1E)}(r)| 0\tilde{s} \rangle
          \nonumber \\
 &-& \frac{2}{\,5\,}\langle 0\tilde{p} |V^{(^3O)}(r)| 0\tilde{p} \rangle
  +\frac{13}{\,15\,}\langle 0\tilde{p} |V^{(^1O)}(r)| 0\tilde{p} \rangle
           \nonumber \\
 &+& \frac{72\sqrt{2}}{75}
 \langle 0\tilde{p}(S_0) |V^{(ALS)}(r)| 0\tilde{p}
  (S_1)\rangle_{\mathcal{I}=1} \ . 
\label{Eq:RelativeMx}
\end{eqnarray}
Here $0\tilde{s}$ and $0\tilde{p}$ denote the relative
radial states of $\Lambda N$ two-body system with the
spin coupling $(\bm{l}+\bm{S}=\bm{\mathcal{I}})$.
Each central potential $V^{(c)}(r)$ is given by sum of the
correponding terms defined by Eqs.(\ref{eq:LN}) - (\ref{eq:YNGc}),
although in the above expression the  $SLS$ spin-orbit
component is formaly included in the
$\langle 0\tilde{p} |V^{(^3O)}(r)| 0\tilde{p} \rangle$ part.
Equation (\ref{Eq:RelativeMx}) shows clearly that, in
addition to the relative $\tilde{s}$-state interactions,
the $\tilde{p}$-state central interactions and the
SLS/ALS interactions contribute in this way to the
$p$-shell hypernuclear doublet splitting. In this respect
it is interesting to compare the expression of
Eq.(\ref{Eq:RelativeMx}) with that for the doublet
splitting in
$^4_{\Lambda}\mbox{H}$
 in the lowest configuration:
\begin{eqnarray}
\lefteqn{ \varDelta E_4(1^+ -0^+_g) } \nonumber \\
&\equiv&  E(^4_{\Lambda}\mbox{H}; 1^+) -  E(^4_{\Lambda}\mbox{H};0^+_g)
     \nonumber \\
&=&\langle 0\tilde{s} |V^{(^3E)}(r)| 0\tilde{s} \rangle
- \langle 0\tilde{s} |V^{(^1E)}(r)| 0\tilde{s} \rangle,
\label{Eq:He4E1_0}
\end{eqnarray}
where only the even-state interactions work.
Also we note that the oscillator size ($b$) for $s$-shell
hypernuclei is implicit and it is different from
the $p$-shell cases.
It should be noted that, in actual calculations, the even- and odd-state combinations of 
$\Lambda N$ spin-dependent interactions are determined carefully within the freedom as prescribed by Eq. (\ref{eq:VevenOdd}), 
{\it i.e.} the contributions of the even- and odd-state components can be different from those given 
by Eq. (\ref{Eq:RelativeMx}) due to dynamical effects such as nuclear deformations 
and/or density dependence of the $\Lambda N$ interaction, 
which are fully taken into account in the HyperAMD calculation using Eq. (\ref{eq:VevenOdd})
as presented in the following section.
%It should be noted that, in the actual calculations, the
%even- and odd-state combinations of $\Lambda N$
%spin-dependent interactions are determined in wide
%flexibility as prescribed by
%Eq.(\ref{eq:VevenOdd}). We also remark that the dynamical
%effects coming from nuclear deformation and density
%dependence of interactions are fully taken into account in the
%HyperAMD calculations as presented in the following section.

\section{Numerical Results and Discussions}
%%%%%%%%%%%%%%%%%%%%%%%%%%
\begin{figure*}
  \begin{center}
    \includegraphics[keepaspectratio=true,width=160mm]{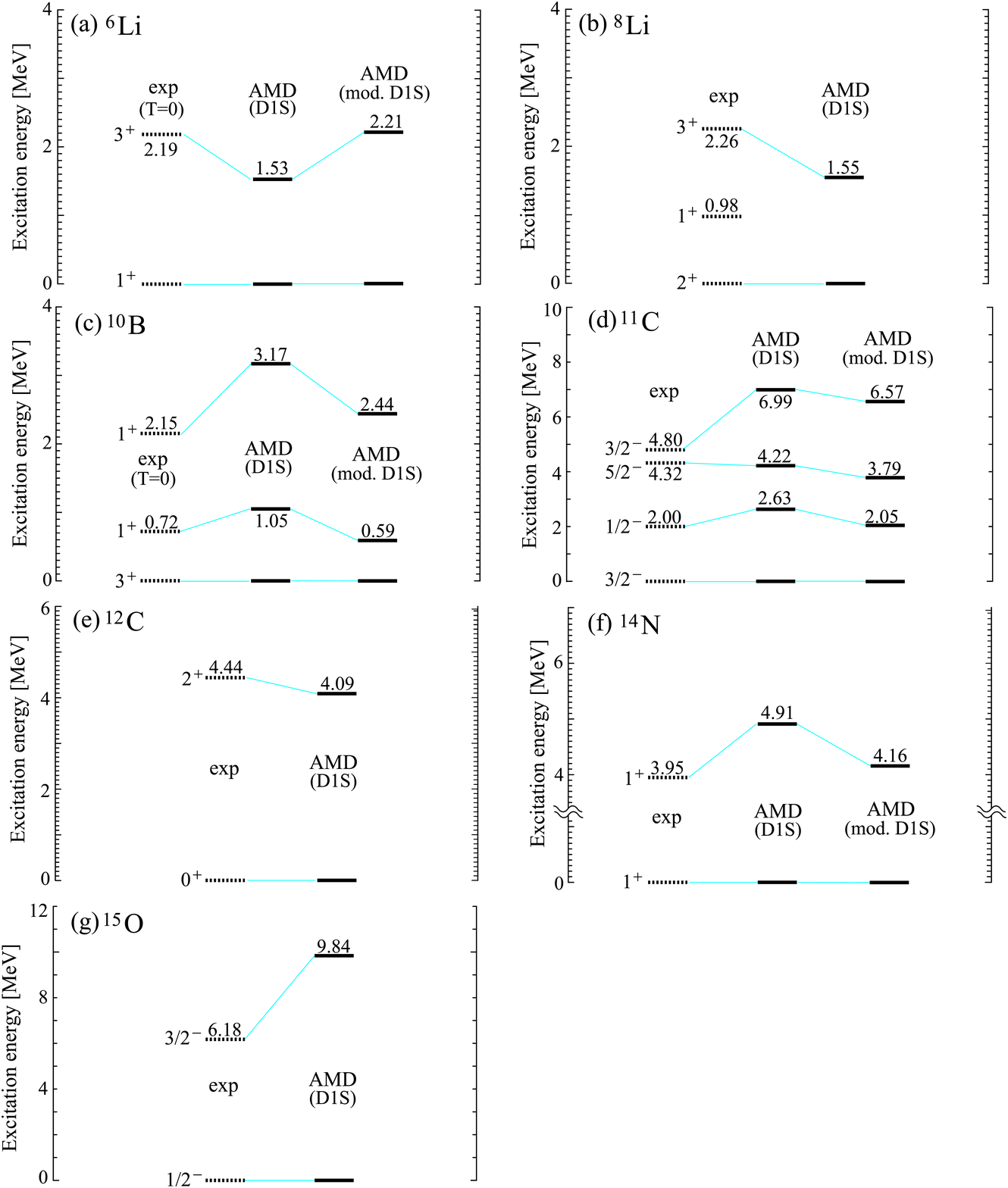}
  \end{center}
  \caption{
Calculated and observed excitation spectra of the core nuclei $^6$Li, $^8$Li, $^{10}$B, $^{12}$C, $^{13}$C, $^{14}$N, and $^{15}$O.
Numbers shown in the spectra are excitation energies [MeV].
In panels (a), (b),(c),(d) and (f),
the calculated results with the Gogny D1S modified so as to reproduce $Ex(^{6}{\rm Li};3^+)$ (mod. D1S) are shown
together with those with the original parameter set of the Gogny D1S (D1S).
The observed data are taken from Ref. \cite{exp_Li6} for $^6$Li, Ref. \cite{exp_C11} for $^{11}$C,
Ref. \cite{exp_C12} for $^{12}$C, Ref. \cite{exp_N14_O15} for $^{14}$N and $^{15}$O,
and Ref. \cite{exp_Li8-B10} for the others.
}
  \label{fig:spectra1-1.eps}
\end{figure*}
%%%%%%%%%%%%%%%%%%%%%%%%%%

%%%%%%%%%%%%%%%%%%%%%%%%%%
\begin{figure*}
  \begin{center}
    \includegraphics[keepaspectratio=true,width=160mm]{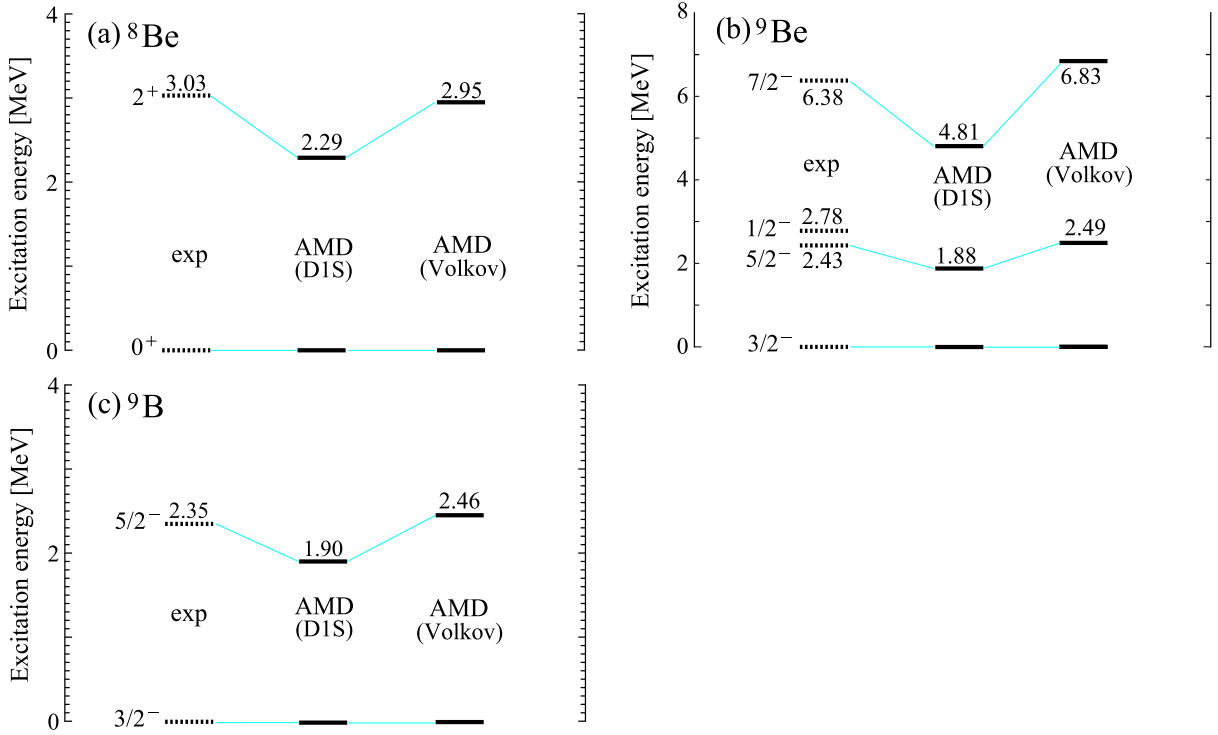}
  \end{center}
  \caption{
Same as Fig.\ref{fig:spectra1-1.eps} but for $^8$Be, $^9$Be and $^9$B.
The calculated results with Volkov No.2 (Volkov) are compared with those with Gogny D1S (D1S).
The observed data are taken Ref. \cite{exp_Li8-B10}.
}
  \label{fig:spectra1-2.eps}
\end{figure*}
%%%%%%%%%%%%%%%%%%%%%%%%%%

In this section, we discuss low-lying level structure affected by $\Lambda N$ spin-spin and spin-orbit interactions.
In Ref. \cite{PRC95.044308(2017)}, the spin orderings of the ground-state doublet are inconsistent with the observations
in $^{10}_\Lambda$B, $^{11}_\Lambda$B, $^{12}_\Lambda$B, $^{12}_\Lambda$C, $^{15}_\Lambda$N, and $^{16}_\Lambda$O,
where the HyperAMD calculations were performed using the $\Lambda N$ central force derived from ESC14+MBE but without the $\Lambda N$ spin-orbit force.
These behaviors would be caused by the properties of the $\Lambda N$ spin-spin force and/or absence of the $\Lambda N$ spin-orbit force.
In this paper, we focus on the spin-spin and spin-orbit splittings
in $^7_\Lambda$Li, $^9_\Lambda$Li, $^9_\Lambda$Be, $^{10}_\Lambda$Be, $^{10}_\Lambda$B, $^{11}_\Lambda$B,
$^{12}_\Lambda$C, $^{13}_\Lambda$C, $^{15}_\Lambda$N, and $^{16}_\Lambda$O.

\subsection{Excitation spectra of core nuclei}
\label{IIIA}

In this study, the excitation spectra of hypernuclei are compared
with those of the core nuclei and observed data of the hypernuclei.
Since the experimental information of hypernuclei is concentrated
in the energy regions near the ground states,
it is quite important to describe the low-lying spectra of the core nuclei properly before the discussion of the hypernuclei.

We perform the antisymmetrized molecular dynamics (AMD) calculation
for $^6$Li, $^8$Li, $^8$Be, $^9$Be, $^9$B, $^{10}$B, $^{11}$C, $^{12}$C, $^{14}$N, and $^{15}$O
using the original parameter set of Gogny D1S
%that is the same effective nuclear force as used
given in Refs. \cite{Gogny1,Gogny2}.  %\cite{PRC94.044310(2016),PRC95.044308(2017)}.
These nuclei have various structure.
For example, in $^{8}$Be and $^{9}$Be,
it is well known that 2$\alpha$ clustering is dominated near the ground states,
whereas the ground state $3/2^-$ of $^{15}$O is well understood
by a neutron one-hole configuration in the shell-model picture.
In $^{12}$C, the ground state has a mixed nature of
the $3\alpha$ cluster and shell-model like structures
\cite{Enyo_1998, Itagaki_2004, Neff_2004, Enyo_2007, Enyo_2012, Suhara_2015}.

First, we discuss the excitation spectra of the nuclei other than $^8$Be, $^9$Be and $^9$B, namely
$^6$Li, $^8$Li, $^9$B, $^{10}$B, $^{11}$C, $^{12}$C, $^{14}$N, and $^{15}$O,
which are shown in Fig. \ref{fig:spectra1-1.eps}.
It is found that the present AMD calculation successfully reproduces the spin-parity of the ground states.
The low-lying states in these nuclei are obtained with correct ordering except for $^{8}$Li.
It is noted that the $1^+$ state of $^{8}$Li is also obtained at 5.27 MeV, which is higher than the $3^+$ state.
In $^{12}$C, since the AMD model can describe both the cluster and shell-model like structures,
we obtain a reasonable value of the excitation energy of the $2^+$ state,
which is sensitive to the mixed nature of the ground state mentioned above.

Though the AMD calculation reproduces the gross features of the nuclei such as the level ordering,
we see deviations of the calculated excitation energies from the experimental data.
For example, in $^6$Li, the excitation energy of $3^+$, $E_x(^6{\rm Li}; {\rm 3}^+)$, is lower
than the experimental data by about 700 keV,
which is quite important because the coupling of a $\Lambda$ particle
to this state generates an experimentally well-known spin doublet $(5/2^+, 7/2^+)$ in $^7_\Lambda$Li.
In the other nuclei, we see the deviation of the excitation energies from the observed data by about several hundred keV or more.
For the quantitative discussions of the hypernuclei, these deviations should be improved as possible.

Here, we try to modify the parameters of Gogny D1S so as to reproduce the $E_x(^6{\rm Li}; {\rm 3}^+)$ value.
The central force of the Gogny D1S is given by the two-range Gaussian form as
\begin{eqnarray}
\nonumber
V^{\rm cent}_{NN} &=& \sum^2_{i=1} \exp \left[ - (r/\beta_i)^2 \right]  \\
     &\ & \times \{ W_i + B_i \hat{P}_\sigma - H_i \hat{P}_\tau -M_i \hat{P}_\sigma \hat{P}_\tau \},
\end{eqnarray}
where $\hat{P}_\sigma$ and $\hat{P}_\tau$ are the spin and isospin exchange operators, respectively.
In general, varying the parameters $B_i$ and $H_i$ controls
the ratio of the $^3 E$ interaction to the $^1 E$ interaction with keeping the sum of $W_i$ and $M_i$.
Since these parameters can be changed depending on each nuclear system,
we modify $B_i$ and $H_i$ for the both ranges $(i = 1, 2)$ 
by multiplying a common factor $\alpha^{NN}$ so as to reproduce the $E_x(^6{\rm Li}; {\rm 3}^+)$ value,
while the $W_i$ and $M_i$ are unchanged from the original values of the Gogny D1S. 
As a result, we find that the $\alpha^{NN} = 1.2$ gives $E_x(^6{\rm Li}; {\rm 3}^+) = 2.21$ MeV,
which is much close to the observed value $E^{\rm exp}_x(^6{\rm Li}; {\rm 3}^+) = 2.19$ MeV.
It is also found that this modification improves the excitation spectra
of $^{10}$B, $^{11}$C, and $^{14}$N as shown in Fig. \ref{fig:spectra1-1.eps} except for the $3/2^-_2$ state of $^{11}$C.
On the other hand, in $^{15}$O, the excitation energy of the $3/2^-$ state is largely different from the observed value.
We consider that this difference mainly comes from the description of neutron single-particle orbits,
because the $3/2^-$ state is generated by a neutron from $0p_{3/2}$ to $0p_{1/2}$ in a naive shell-model picture.

In Fig. \ref{fig:spectra1-2.eps}, we show the excitation spectra of $^8$Be, $^9$Be and $^9$B,
which are characterized by having the $2\alpha$, $2\alpha + n$ and $2\alpha + p$ cluster structure, respectively.
It is found that the present calculation with Gogny D1S reproduces the level ordering except for the $1/2^-$ state in $^{9}$Be and $^9$B.
However, the excitation energies are slightly lower than the experimental data in $^8$Be, $^9$Be and $^9$B.
Since they have the well-developed 2$\alpha$ cluster structure,
it is considered that the difference of the excitation energies shows the overestimation of the momentum of inertia of the the 2$\alpha$ clustering.
In general, a density-dependent interaction tends to overestimate the nuclear clustering,
because it works as a repulsive force when the nuclear density is increased in cluster states.
Since the degree of the 2$\alpha$ clustering as well as the size of the $\alpha$ particle affect
the momentum of inertia, the difference of the excitation energies in Fig. \ref{fig:spectra1-2.eps} would be caused
by the density-dependent interaction of Gogny D1S.
Instead of the Gogny D1S, we use the Volkov No. 2 force \cite{Volkov2} together with the spin-orbit force
of the G3RS interaction \cite{G3RS} for $^8$Be, $^9$Be and $^9$B (see Fig. \ref{fig:spectra1-2.eps}).
The Volkov No. 2 is one of the effective nuclear forces often used in the structure studies on $\alpha$ cluster structure, %which is given by
which does not have the density-dependent term as given by
\begin{eqnarray}
\nonumber
V^{\rm cent}_{NN} &=& \sum^2_{i=1} v_i \exp \left[ - (r/\beta_i)^2 \right] \\
&\ & \times \{ (1.0 - m) + b P_\sigma - h P_\tau - m P_\sigma P_\tau \}.
\end{eqnarray}
Here, the parameters are fixed to be $m=0.60$ and $b=h=0.125$.
The strength of the spin-orbit force of the G3RS is fixed to be 1600 MeV.
In Fig. \ref{fig:spectra1-2.eps}(a), in $^8$Be,
we see that the excitation energy of the 2$^+$ state becomes 2.95 MeV with Volkov No. 2,
which is close to the experimental value ($E^{\rm exp}_x (^{8}{\rm Be}; 2^+) = 3.03$ MeV).
The energy spectra of $^9$Be and $^9$B are also improved as shown in Fig. \ref{fig:spectra1-2.eps}(b)-(c).

We also give a comment on the $1/2^-$ state in $^9$Be and $^9$B, which are missing in Fig. \ref{fig:spectra1-2.eps}(b)-(c).
This is caused by the procedure of the variational calculation.
In $1/2^-$ state of $^9$Be, the last neutron occupies the $p_{1/2}$ orbit in the $2\alpha + n$ cluster model picture.
In the present calculation, the nucleon configuration is determined by the energy variation performed before the angular momentum projection.
As a result, the last neutron mainly occupies the $p_{3/2}$ orbit than the $p_{1/2}$ orbit.
In $^9$B, which is the mirror nucleus of $^9$Be, the $1/2^-$ state does not appear
because the last proton occupies the $p_{3/2}$ orbit as the results of the energy variation.

From the above results, we use the modified version of the Gogny D1S as the effective nuclear interaction
in the HyperAMD calculation for $^{7}_\Lambda$Li,
$^{11}_\Lambda$B, $^{12}_\Lambda$C, and $^{15}_\Lambda$N,
whereas the original parameter set of it is employed for $^{9}_\Lambda$Li, $^{13}_\Lambda$C, and $^{16}_\Lambda$O.
In $^{9}_\Lambda$Be, $^{10}_\Lambda$Be and $^{10}_\Lambda$B,
the Volkov No.2 force is used together with the spin-orbit force of the G3RS instead of Gogny D1S.

%%%%%%%%%%%%%%%%%%%%%%%%%%
\begin{figure*}
  \begin{center}
    \includegraphics[keepaspectratio=true,width=120mm]{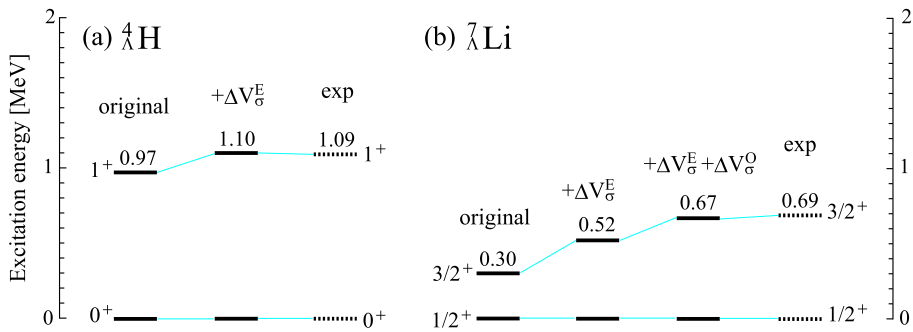}
  \end{center}
  \caption{
(a) Comparison of calculated (solid) and observed (dotted) excitation spectra of $^{4}_\Lambda$H.
Calculated spectrum with the original parameter set of ESC14+MBE is shown in the left (original),
while that using ESC14+MBE with tuning the $\Lambda N$ even-state spin-spin force is shown in the middle ($+ \Delta V^{\rm E}_{\sigma}$).
Numbers shown in the spectra are excitation energies [MeV].
(b) Same as (a) but for $^{7}_\Lambda$Li.
Calculated spectrum using ESC14+MBE with the tuning in the $\Lambda N$ even-state (even and odd state) spin-spin force(s) are shown in the 2nd (3rd) from the left,
while that without tuning is displayed in the left.
Experimental data are taken from Refs. \cite{Bamberger1971,Bedjidian1976,May1983,Tamura2000}.
}
  \label{fig:tuning.eps}
\end{figure*}
%%%%%%%%%%%%%%%%%%%%%%%%%%

\subsection{Tuning of spin-dependent $\Lambda N$ interaction}
\label{IIIB}

Let us move to the discussions on the hypernuclei.
In this section, we tune the $\Lambda N$ spin-spin and spin-orbit interactions in light $\Lambda$ hypernuclei.
First, we focus on the spin doublet $(0^+, 1^+)$ of $^4_\Lambda$H
generated by the coupling of a $\Lambda$ particle to the $1/2^+$ state of $^3$H,
where the even-state spin-spin force $v^E_\sigma$ dominantly acts between each nucleon and the $\Lambda$ particle in $s$ orbit.
In Fig. \ref{fig:tuning.eps}(a), we show the calculated result of $^4_\Lambda$H with the original parameter sets
of Gogny D1S and ESC14+MBE as the effective nuclear and $\Lambda N$ interactions, respectively.
Here we use the $k_F$ value which reproduces the $B_\Lambda^{\rm exp}$ value of $^{4}_\Lambda$H,
namely  $k_F = 0.95$ fm$^{-1}$ ($k_F = 0.93$ fm$^{-1}$) giving $B_\Lambda = 2.09$ MeV ($B_\Lambda = 2.14$ MeV) with (without) tuning.
It is seen that the excitation energy of the $1^+$ state is slightly underestimated (0.97 MeV)
in comparison with the observed value ($E^{\rm exp}_x (^{4}_\Lambda {\rm H}; 1^+) =1.09 \pm 0.02$ MeV).
Here we tune the even-state spin-spin force $v_\sigma^E$ by adding a correction term $\Delta V^E$ to the 2nd range as
\begin{eqnarray}
 \Delta V^E \exp \left[ -(r/\beta_2)^2 \right] \vec{\sigma} \cdot \vec{\sigma} P(E).
\end{eqnarray}
As a result, the experimental value $E^{\rm exp}_x (^{4}_\Lambda {\rm H}; 1^+)$ is reproduced
by using $\Delta V^E = 1.2 $ MeV as shown in the middle of Fig. \ref{fig:tuning.eps}(a).

To tune the odd-state spin-spin force $v_\sigma^O$, we focus on the excitation spectra of $^7_\Lambda$Li,
where the spin doublets $(1/2^+, 3/2^+)$ and $(5/2^+, 7/2^+)$ are observed corresponding to the $1^+$ and $3^+$ states of $^6$Li, respectively.
One can consider that $^6$Li has the cluster structure composed of the spin-saturated $\alpha$ particle and the $d$ cluster with spin 1,
the relative angular momentum between $\alpha$ and $d$ is zero $l=0$ in the ground state $1^+$,
whereas the $3^+$ state is generated by the coupling of $l=2$ and spin 1 of $d$.
In $^7_\Lambda$Li, the spin-spin forces in both even and odd parity states contribute to the $(1/2^+, 3/2^+)$ and $(5/2^+, 7/2^+)$.
It is noted that the $\Lambda N$ spin-orbit force acts only in the $(5/2^+, 7/2^+)$ state,
because the $\Lambda$ particle mainly occupies the $s$ orbit in $^7_\Lambda$Li.
Here, we calculate the lowest doublet $(1/2^+, 3/2^+)$ using the ESC14+MBE with $\Delta V^E = 1.2 $ MeV as shown in Fig. \ref{fig:tuning.eps}(b),
where the result with the original parameter set of ESC14+MBE is also shown.
It is seen that the excitation energy of $3/2^+$ is increased by adding $\Delta V^E = 1.2$ MeV (0.52 MeV),
but still lower than the experimental value (0.69 MeV).
In the same manner as the even-state spin-spin force, we add the correction term $\Delta V^O $ as
\begin{eqnarray}
 \Delta V^O \exp \left[ -(r/\beta_2)^2 \right] \vec{\sigma} \cdot \vec{\sigma} P(O),
\end{eqnarray}
to the odd-state spin-spin force and determine $\Delta V^O = 30.0$ MeV for reproducing $E_x (^7_\Lambda{\rm Li}: 3/2^+)$
(see the 3rd from the left in Fig. \ref{fig:tuning.eps}(b)).
Since $\Delta V^E$ and $\Delta V^O$ are $k_F$ independent,
one can renormalize these corrections into $a^{(c)}_i$ in Eq. (\ref{eq:YNGe}) using the relation among Eqs. (\ref{eq:LN}) - (\ref{eq:VevenOdd}).
The corrected parameters $a^{(c)}_i$ for Eq. (\ref{eq:YNGe}) are shown in parenthesis in Tab. \ref{Tab:central}.

In addition to the spin-spin force, we also modify the $\Lambda N$ spin-orbit force so as to reproduce the energy splitting between $3/2^+$ and $5/2^+$ of $^9_\Lambda$Be.
Since $^8$Be has well pronounced 2$\alpha$ cluster structure,
the 1st excited state $2^+$ is mainly generated by the relative angular momentum $l = 2$ between 2$\alpha$.
If a $\Lambda$ particle is injected to this state,
the energy splitting between $3/2^+$ and $5/2^+$ is mainly caused by the $\Lambda N$ spin-orbit force.
In $^{9}_\Lambda$Be, using the original SLS and ALS force, the energy splitting between the $3/2^+$ and $5/2^+$ states is 150 keV,
which is much larger than the experimental value (40 keV).
It is well known that the spin-orbit splitting by the $\Lambda N$ interaction is quite small compared with that by nuclear force,
which is due to the large cancellation between SLS and ALS forces.
Therefore, we strengthen the ALS by multiplying a factor $\alpha_{\rm ALS}$ to make the cancellation larger with keeping the strength of the SLS.
It is found that the energy spacing between $3/2^+$ and $5/2^+$ becomes smaller as the $\alpha_{\rm ALS}$ increases.
As a result, we determine $\alpha_{\rm ALS} = 1.9$, which gives the splitting energy of 60 keV.
%Using $\alpha_{\rm ALS} = 1.9$ together with $\Delta V^E = 1.2 $ MeV and $\Delta V^O = 30.0$ MeV,
%the energy spacing of the $(5/2^+, 7/2^+)$ doublet in $^7_\Lambda$Li becomes 0.48 MeV,
%which is much close to the observed value (0.47 MeV) (see also Fig. \ref{fig:spectra2.eps}(a)).

Hereafter we use the $\Lambda N$ central force of ESC14+MBE with both $\Delta V^E = 1.2 $ MeV and $\Delta V^O = 30.0$ MeV
as well as the $\Lambda N$ spin-orbit force with $\alpha_{\rm ALS} = 1.9$.

\subsection{Ground-state spin of the hypernuclei}

%%%%%%%%%%%%%%%%%%%%%%%%%%
\begin{figure*}
  \begin{center}
    \includegraphics[keepaspectratio=true,width=150mm]{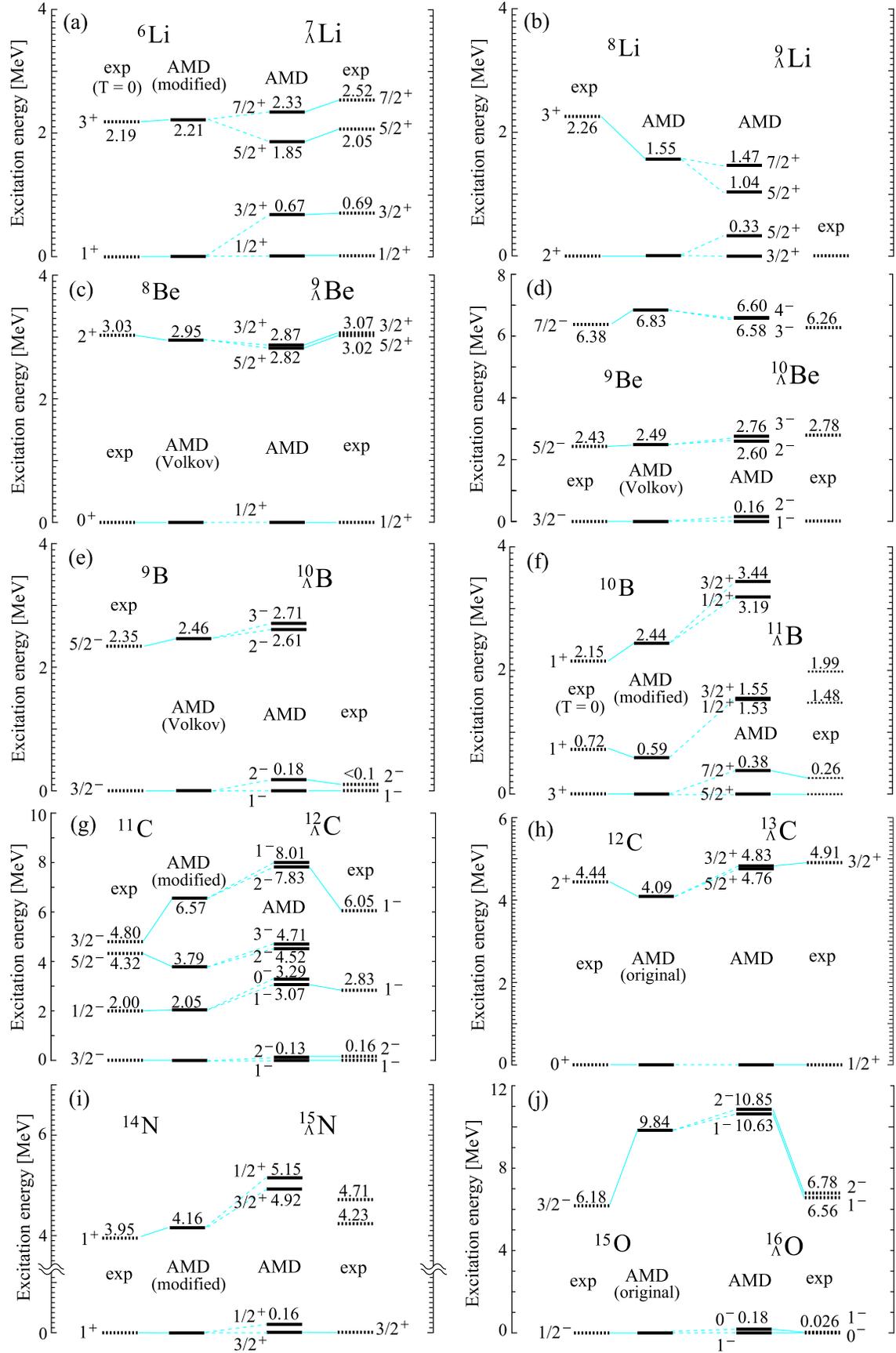}
  \end{center}
  \caption{
Calculated (solid) and observed (dotted) excitation spectra.
Numbers shown in the spectra are excitation energies [MeV].
Observed data of the excitation spectra are also shown for
$^{7}_\Lambda$Li \cite{May1983,Tamura2000},
$^{9}_\Lambda$Be \cite{Akikawa2002},
$^{10}_\Lambda$Be \cite{Gogami2016},
$^{10}_\Lambda$B \cite{Tamura2005,Chrien1990},
$^{11}_\Lambda$B \cite{Miura2005},
$^{12}_\Lambda$C \cite{Hosomi2015},
$^{13}_\Lambda$C \cite{Kohri2002,PPNP57.564(2006)},
$^{15}_\Lambda$N \cite{Tamura2005}, and
$^{16}_\Lambda$O \cite{Ukai2004}.
}
  \label{fig:spectra2.eps}
\end{figure*}
%%%%%%%%%%%%%%%%%%%%%%%%%%

Using the $\Lambda N$ interaction tuned in Sec. \ref{IIIB}, we apply the HyperAMD to the $p$ shell hypernuclei, $^{7}_\Lambda$Li, $^{9}_\Lambda$Li, $^{9}_\Lambda$Be, $^{10}_\Lambda$Be, $^{10}_\Lambda$B, $^{11}_\Lambda$B, $^{12}_\Lambda$C, $^{13}_\Lambda$C, $^{15}_\Lambda$N, and $^{16}_\Lambda$O, which are shown in Fig. \ref{fig:spectra2.eps}.

Since the YNG interaction depends on the nuclear density through the Fermi momentum $k_F$, it is necessary to select the $k_F$ properly in each system.
%In this paper, we use the same $k_F$ values as in Ref. \cite{PRC95.044308(2017)}, determined under the ADA, except for $^{7}_\Lambda$Li, $^{9}_\Lambda$Li, $^{9}_\Lambda$Be and $^{10}_\Lambda$Be.
In this paper, except for $^{7}_\Lambda$Li, $^{9}_\Lambda$Li, $^{9}_\Lambda$Be, $^{10}_\Lambda$Be and $^{10}_\Lambda$B,
we use the same $k_F$ values as in Ref. \cite{PRC95.044308(2017)} determined under the ADA,
where the $k_F$ values are obtained from the densities of the nucleons ($\rho_N$) and $\Lambda$ particle ($\rho_\Lambda$), as
\begin{eqnarray}
k_F = \left( \frac{3\pi^2 \langle \rho \rangle}{2} \right), \ \langle \rho \rangle = \int d^3r \rho_N (\bf{r}) \rho_\Lambda (\bf{r}).
\label{Eq:ADA}
\end{eqnarray}
The $k_F$ values used are listed in Tab. \ref{Tab:hyper}, where the common $k_F$ value is used for the ground and excited states in each hypernucleus.
The $\Lambda$ binding energy $B_\Lambda$ is also shown in Tab. \ref{Tab:hyper}, which is defined by the energy gain of a hypernuclear state from the corresponding state in the core nucleus.
It is found that the $B_\Lambda$ values  are increased compared with those in Ref. \cite{PRC95.044308(2017)}, and thus are larger than the observed values ($B_\Lambda^{\rm exp}$).
For example, in $^{12}_\Lambda$C, the $B_\Lambda$ is 12.28 MeV in Tab. \ref{Tab:hyper},
whereas $B_\Lambda = 11.0$ MeV in Ref. \cite{PRC95.044308(2017)}.
This is mainly due to the inclusion of the $\Lambda N$ spin-orbit force.
Since the ADA given by Eq.(\ref{Eq:ADA}) is one of the methods to obtain reasonable values of $k_F$ from the structure calculation,
there is a room to modify the ADA to reproduce $B_\Lambda^{\rm exp}$.
For example, in Ref. \cite{PRC95.044308(2017)}, the authors (M.I and Y.Y) introduced a small correction parameter in Eq. (\ref{Eq:ADA}) for the $p$-states of several $\Lambda$ hypernuclei.
In the same way, it is possible to modify the ADA using a small correction parameter in the present calculation.
In $^{7}_\Lambda$Li, $^{9}_\Lambda$Li, $^{9}_\Lambda$Be and $^{10}_\Lambda$Be,
the $k_F$ values are determined so as to reproduce the observed values of the $B_\Lambda$ in the ground states.
This is because the HyperAMD with the ADA treatment largely overestimates the $B_\Lambda$ values in the light hypernuclei ($A < 10$)
and/or the hypernuclei with $\alpha$ clustering as pointed out in Ref. \cite{PRC95.044308(2017)}.

Let us discuss the spin-parity of the ground states of the hypernuclei.
In Fig. \ref{fig:spectra2.eps}, the excitation spectra of the hypernuclei are compared with those of the core nuclei and experimental data.
It shows that the calculation successfully reproduces the level ordering of the ground-state doublets
of $^{7}_\Lambda$Li, $^{10}_\Lambda$B, $^{11}_\Lambda$B, $^{12}_\Lambda$C, and $^{15}_\Lambda$N,
in which incorrect spin was obtained for the ground states in Ref. \cite{PRC95.044308(2017)}.
The difference of the calculations between Ref. \cite{PRC95.044308(2017)} and this paper is the $\Lambda N$ interaction,
namely tuning of the spin-spin force and inclusion of the spin-orbit force, which improves the incorrect ordering of the spin doublet partners.
In $^{16}_\Lambda$O, the degenerated ground state doublet is obtained in the calculation, though the ground-state spin is different with the observed one.
Thus the $\Lambda N$ spin-spin and spin-orbit interactions are essential to describe the ground state spin properly.

We predict the ground-state spin and parity of $^9_\Lambda$Li and $^{10}_\Lambda$Be, which are not measured by experiments.
In $^9_\Lambda$Li, Figure \ref{fig:spectra2.eps} (b) shows that the $3/2^+$ state is lower than the $5/2^+$ state in the ground-state doublet,
which is opposite to the prediction in Ref. \cite{PRC95.044308(2017)} where the $5/2^+$ state is predicted to be the ground state.
Similarly, in Fig. \ref{fig:spectra2.eps}(d), the $1^-$ states is predicted as the ground states of $^{10}_\Lambda$Be,
whereas the $2^-$ is the lowest in Ref. \cite{PRC95.044308(2017)}.
From the above discussion, we consider that the ground-state spin predicted by the present work is more reliable than that in Ref. \cite{PRC95.044308(2017)}.
Furthermore, it is also found that the above tuning of ESC14+MBE reproduces the observed ground-state spin of $^{19}_\Lambda$F. The detailed result of $^{19}_\Lambda$F will be reported in a forthcoming paper.

\subsection{Comparison of the excitation spectra with the experimental data}

In this section, we discuss not only the ground states but also the excited states of each hypernuclei in detail in comparison with the observed data.

In $^{7}_\Lambda$Li, it is found that both of the splitting and excitation energies
of the $(5/2^+, 7/2^+)$ doublet are almost consistent with the experimental data.
In particular, the splitting energy of this doublet is important to test the tuned $\Lambda N$ interaction,
where the $\Lambda N$ spin-orbit force also affects in addition to the even and odd state spin-spin force.
This is because the core state $3^+$ of $^6$Li is generated by the coupling of the relative angular momentum $l=2$
between $\alpha$ and $d$ and the spin 1 of the $d$ in a naive $\alpha$ + $d$ cluster model picture.
In Fig. \ref{fig:spectra2.eps}, the splitting energy of the $(5/2^+, 7/2^+)$ doublet is 0.48 MeV, which is almost the same as the observed value (0.47 MeV), whereas that is 0.33 MeV if the original parameter set of the $\Lambda N$ spin-orbit interaction is used.
Thus the energy splitting of the $(5/2^+, 7/2^+)$ doublet is consistently obtained by the tuning of the $\Lambda N$ spin-orbit interaction in $^{9}_\Lambda$Be in Sec. \ref{IIIB}.

In the other hypernuclei, the excitation spectra are successfully described using the $\Lambda N$ interaction tuned in Sec. \ref{IIIB}.
In $^{10}_\Lambda$B, the energy splitting of the ground-state doublet is small (0.18 MeV).
In $^{13}_\Lambda$C, the excitation energy of the $3/2^+$ is consistent with that measured by the $\gamma$-ray spectroscopy experiment.
In $^{10}_\Lambda$Be, the excitation energies of the $(2^-_2, 3^-_1)$ and $(3^-_2, 4^-)$ doublets are consistent with the excited states observed at 2.78 MeV and 6.26 MeV with unknown spin-parity.
In $^{11}_\Lambda$B, the present calculation almost reproduces the energy spacing of the observed ground-state doublet.
It is found that the contribution from the odd-state spin-spin interaction is quite small in the ground-state doublet of $^{11}_{\Lambda}$B,
and thus the even-parity spin-spin and spin-orbit interactions are important to reproduce the energy spacing.
Furthermore, one of the observed excited states at 1.48 MeV in $^{11}_\Lambda$B is close to the calculated $(1/2^+, 3/2^+)$ doublet.
In $^{12}_\Lambda$C, the present calculation nicely reproduces the ground-state doublet $(1^-_1, 2^-_1)$. In addition, the excitation energy of the $1^-_2$ states at 3.07 MeV is also close to the observed value (2.83 MeV).

In the several hypernuclei, the excitation energies of the doublets are higher than those of the observed states,
which is mainly due to the deviation of the core excitation energies from the experiments.
For example, in the $1^-_3$ state of $^{12}_\Lambda$C, the calculated excitation energy ($E_x (^{12}{\rm C};1^-_3) = 8.01$ MeV) is higher by about 2 MeV than the observed value.
Note that the excitation energy of the core state $3/2^-_2$ is overestimated by about 1.8 MeV.
By subtracting the energy shift of 1.8 MeV from the excitation energy, the $E_x (^{12}{\rm C};1^-_3)$ becomes much close to the observed value (6.05 MeV).
The same thing occurs in $^{16}_\Lambda$O: the excited $(1^-_2, 2^-_1)$ doublet is rather high in excitation energy,
whereas the splitting energy (0.22 MeV) is consistent with the observation.
This is mainly due to the overestimation of the excitation energy of the core state $3/2^-_2$ in $^{15}$O,
related to the description of the single-particle orbits as mentioned in Sec. \ref{IIIB}.
Thus, in these cases, the discrepancy of the excitation energies is attributed to those of the core states.
Concerning this fact, in $^9_\Lambda$Li,
the excitation energy of the $(5/2^+, 7/2^+)$ doublet would be higher by about 0.7 MeV,
because the core state $3^+$ locates lower than the observed stata by 0.71 MeV in $^8$Li.

Let us discuss the excited states of $^{15}_\Lambda$N.
In $^{15}_\Lambda$N, the splitting energy (0.47 MeV) of the $(1/2^+_2, 3/2^+_2)$ doublet is
almost the same as the energy spacing (0.48 MeV) of the observed excited states at 4.23 MeV and 4.71 MeV.
Therefore these observed states are considered to be $(1/2^+_2, 3/2^+_2)$ doublet.
However, the excitation energies are higher by about 0.7 MeV than those of the observed states.
Since the deviation of the core state $1^+_2$ state in $^{14}$N from the experiment is only about 0.2 MeV in excitation energy,
it is considered that the present calculation slightly overestimates the energy shift by a $\Lambda$ particle.
In Fig. \ref{fig:spectra2.eps}, we see shifts of the excitation energy in the other hypernuclei, which are discussed in the next Section.

We give comments on the tensor-force contributions as well as $\Lambda N$-$\Sigma N$ coupling
effects in $\Lambda$ hypernuclei. 
It is revealed that the tensor components affect splitting energies of spin-doublets of $\Lambda$
hypernuclei \cite{ AnnPhys(N.Y.).63.53(1971), AnnPhys(N.Y.)116.167(1978), PhysRevC.31.499(1985)}.
It is also pointed out that the $\Lambda N$-$\Sigma N$ coupling contributes to excitation energies
\cite{NuclPhysA.754.48c(2005),NuclPhysA.804.84(2008),RevModPhys.88.035004(2016)}.
Recently, the $\Lambda N$ interactions derived from chiral effective field theory have been developed and 
are used in several hypernuclear structure calculations such as the no-core shell model calculations \cite{PhysRevLett.117.182501(2016),PhysLettB779.336(2018)}. 
For example, in Ref. \cite{PhysRevLett.117.182501(2016)}, the competition between the $\Lambda N$-$\Sigma N$ coupling and $YNN$ three-body force is also investigated in $p$-shell $\Lambda$ hypernuclei. 
The $\Lambda N$-$\Sigma N$ coupling affects not only the excitation energies but the bound-state formation of hypernuclei. 
In Ref. \cite{ PhysRevLett.89.142504(2002)}, it is showed that the tensor components of the $\Lambda N$-$\Sigma N$ coupling plays an essential role to make light hypernuclei bound. 

In the ESC interaction model, the important roles are played by the $\Lambda N$-$\Sigma N$
coupling interactions composed of central and tensor terms. In the $G$-matrix approach, 
high-momentum components of these interactions are renormalized into the $\Lambda N$-$\Lambda N$
central interactions. The residual $\Lambda N$-$\Lambda N$ and $\Lambda N$-$\Sigma N$
interactions also composed of central and tensor terms are not taken into account in this work,
considering that these low-momentum tensor interactions are usually of minor contributions 
for the spectra of $\Lambda$ hypernuclei.
%Similar statements can be given for the residual $\Lambda N$-$\Sigma N$ coupling interactions.
Anyway, it is our future subject to take into account $\Lambda N$-$\Sigma N$ and tensor
couplings explicitly in our AMD treatments. 

Here, let us reveal the importance of the renormalized parts, which is implicit in our results:
We have tried to derive the limited YNG interaction from the ESC+MBE model by switching off all tensor forces
in $\Lambda N$-$\Lambda N$ and $\Lambda N$-$\Sigma N$ channels, and applied it to perform 
calculations for $^4_\Lambda$H, $^7_\Lambda$Li, and $^9_\Lambda$Be. Then, the differences
from the above results demonstrate the tensor-force contributions.
First of all, these hypernuclei are found to be unbound when the tensor-force 
contributions are switched off in deriving the limited YNG interaction.
Here, the dominant contribution to the tensor-force renormalization comes from
the $\Lambda N$-$\Sigma N$ $SD$-coupling tensor term.
%which is essential for bound-state formations of $\Lambda$ hypernuclei.
This shows an essential role of the $\Lambda N$-$\Sigma N$ tensor coupling for 
bound-state formation of $\Lambda$ hypernuclei.
It is also found that switching off the tensor contributions changes the spin-spin splitting
in the excitation spectra dramatically, while the spin-orbit splittings are not affected
so significantly.
In $^4_\Lambda$H, the excitation energy of the $1^+$ state from the ground $0^+$ state, 
$E_x(^4_\Lambda{\rm H};1^+)$, is shifted up and becomes $E_x(^4_\Lambda{\rm H};1^+) = 2.05$ MeV
with the tensor contribution being switched off, which is compared 
%with switching off the tensor contribution, compared 
to the experimental value $E_x^{\rm exp}(^4_\Lambda{\rm H};1^+) = 1.09$ MeV.
This is mainly due to the reduction of the $^3 E$ attraction dominated in the $1^+$ state
by switching off the tensor components, whereas the singlet-even ($^1 E$) force with
no tensor component is important in the ground state $0^+$.
In $^7_\Lambda$Li, the excitation energy of the $3/2^+$ state becomes
$E_x (^7_\Lambda{\rm Li};3/2^+) = 1.53$ MeV, which is much higher than the experimental value
$E_x^{\rm exp}(^7_\Lambda{\rm Li};3/2^+) = 0.69$ MeV.
Thus, it is demonstrated that the bare tensor forces in the ESC model affect significantly
hypernuclear bound-state formations and spin-doublet splittings. 
We emphasize that this mechanism of taking account of tensor-force effect is sufficiently represented
as the effective central attraction under the $G$-matrix approximation. 

\begin{figure}
  \begin{center}
    \includegraphics[keepaspectratio=true,width=86mm]{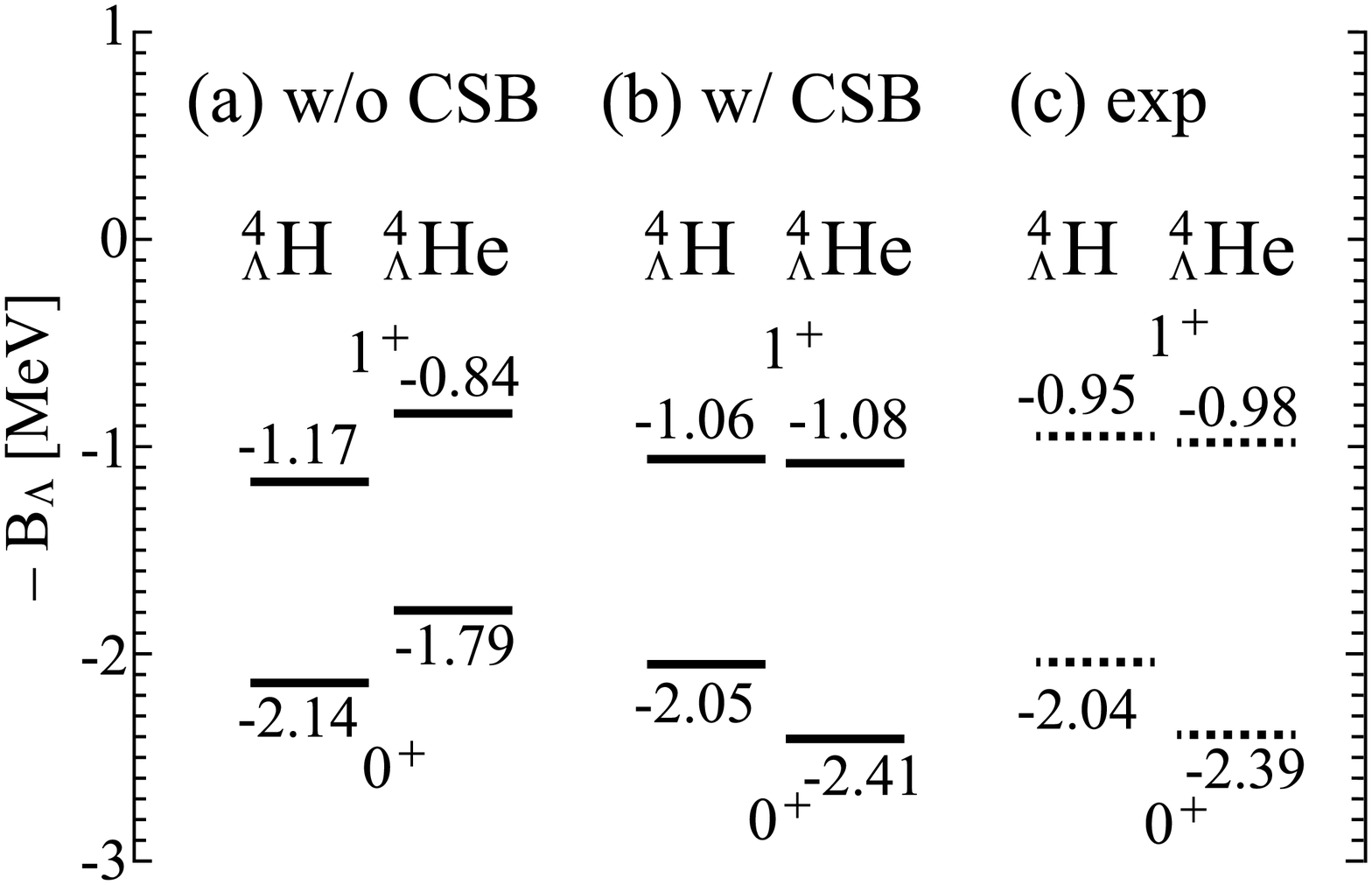}
  \end{center}
  \caption{
            (a) Calculated energy spectra of $^4_\Lambda$H and $^4_\Lambda$He using ESC+MBE with $k_F = 0.92$ fm$^{-1}$. (b) Same as (a), but with the phenomenological charge symmetry breaking (CSB) force explained in text. (c) Experimental data taken from Refs. \cite{BL_Be9L,PhysRevLett.115.222501(2015)}.
          }
  \label{fig:CSB.eps}
\end{figure}

Next, we discuss the charge dependence of the $\Lambda N$ interaction.
In the $A=4$ hypernuclei, the $B_\Lambda$ values of the ground states are known to be different:
$B_\Lambda = 2.04 \pm 0.04$ MeV in $^4_\Lambda$H and $B_\Lambda = 2.39 \pm 0.03$ MeV in $^4_\Lambda$He \cite{BL_Be9L}.
Recently, in $^4_\Lambda$He, the excitation energy of the $1^+$ is precisely measured by the $\gamma$-ray spectroscopy experiment \cite{PhysRevLett.115.222501(2015)}.
The reported value is $E_x(^4_\Lambda{\rm He};1^+) = 1.406 \pm 0.004$ MeV,
which is much larger than $E_x(^4_\Lambda{\rm H};1^+) = 1.09 \pm 0.02$ MeV.
These facts indicate that the $\Lambda N$ interaction has the charge dependence.
Since the $^4_\Lambda$H and $^4_\Lambda$He are the mirror hypernuclei, this fact is referred to the charge symmetry breaking (CSB).
One can expect that the CSB affects the spin-doublet splitting as well as the $\Lambda$ binding energies ($B_\Lambda$).
In this study, we evaluate the CSB effects using the phenomenological CSB force, which is given by the same functional form as in Ref.\cite{NuclPhysA.835.215(2010)}, namely,
\begin{eqnarray}
v^{\rm CSB}(r) = -\frac{\tau_z}{2}
\Big[ \frac{1 + P_x}{2} (v_0^{\rm E,CSB} + \vec{\sigma} \cdot \vec{\sigma} v_\sigma^{\rm E,CSB} ) e^{-r^2} \nonumber \\
+ \frac{1 - P_x}{2} (v_0^{\rm O,CSB} + \vec{\sigma} \cdot \vec{\sigma} v_\sigma^{\rm O,CSB} ) e^{-r^2}
\Big].
\nonumber
\label{Eq:CSB}
\end{eqnarray}
As shown in Fig. \ref{fig:CSB.eps}, we determine the parameters of even-parity part $v_0^{\rm E,CSB}$ and $v_\sigma^{\rm E,CSB}$
so as to reproduce the experimental data of $^4_\Lambda$H and $^4_\Lambda$He,
where the odd-parity parts ($v_0^{\rm O,CSB}$ and $v_\sigma^{\rm O,CSB}$) are zero.
The resulting values are $v_0^{\rm E,CSB} = 10.0$ MeV and $v_\sigma^{\rm E,CSB} = 1.5$ MeV.
We apply this CSB force to several $N \ne Z$ hypernuclei, $^{9}_\Lambda$Li, $^{10}_\Lambda$Be and $^{10}_\Lambda$B.
It is found that the $B_\Lambda$ values are reduced by 0.27 MeV and 0.12 MeV in $^{9}_\Lambda$Li and $^{10}_\Lambda$Be, respectively,
while $B_\Lambda$ is increased by 0.12 MeV in $^{10}_\Lambda$B, depending on the balance of the proton and neutron numbers.
On the other hand, the changes of the excitation energies are about 40 keV at maximum.
Therefore the CSB indicated by the $^4_\Lambda$H and $^4_\Lambda$He data does not affect
the spin-doublet splitting of the $p$-shell hypernuclei significantly.

Finally, we also give a comment on the excitation spectra calculated with the NSC97f force
that is characterized by the strong repulsion of the odd-state interaction,
whereas that is weakly repulsive in the ESC14 model.
It is noted that the central and tensor forces of the $\Lambda N$-$\Sigma N$ coupling are renormalized in NSC97f by the $G$-matrix calculation.
It was pointed out that the central force of NSC97f gives the better agreement
of the spin doublet splitting energies in several $p$-shell hypernuclei \cite{Yam2010},
while its spin-orbit force gives larger splitting energy.
Using the central force of NSC97f, the splitting energies of the ground-state doublets are calculated as 1.11 MeV and 0.60 MeV in $^4_\Lambda$H ($0^+$ and $1^+$) and $^{7}_\Lambda$Li ($1/2^+$ and $3/2^+$), respectively, which are almost consistent with the observed values.
However, it is necessary to tune the spin-orbit force.
In $^9_\Lambda$Be, the splitting energy of the $(3/2^+, 5/2^+)$ doublet is
500 keV using the original parameter set of the spin-orbit force of NSC97f, which is much larger than the observed value (40 keV) and that with the original parameter set of ESC14+MBE.
Therefore, we introduce a factor $\alpha_{\rm ALS}$ multiplied to the ALS force to control the cancellation between the SLS and ALS forces, in the same manner as in Sec. \ref{IIIB}.
It is found that $\alpha_{ALS} = 5.5 $ gives a reasonable value of the splitting energy (50 keV) in $^{9}_\Lambda$Be,
which is much larger than that determined with ESC14+MBE ($\alpha_{ALS} = 1.9 $).
This corresponds to the larger splitting energy with the original spin-orbit force of NSC97f.
Using both the central and above spin-orbit forces,
it is found that the NSC97f interaction gives similar results of the excitation spectra
of the $p$-shell hypernuclei to those using the ESC14+MBE.

%%%%%%%%%%%%%%%%%%%%%%%%%%
%%%%%%%%%%%%%%%%%%%%%%%%%%

%%%%%%%%%%%%%%%%%%%%%%%%
\begin{table*}
  \caption{
Calculated total ($E$), excitation ($E_x$), and $\Lambda$ binding ($B_\Lambda$) energies for each state in the hypernuclei in unit of MeV.
Experimental values of the $\Lambda$ binding energy $B^{\rm exp}_\Lambda$ [MeV] in the ground states are also shown.
In $^{16}_\Lambda$O, the $B^{\rm exp}_\Lambda$ value with \dag  is shifted deeper by 0.54 MeV from that measured by the $(\pi^+, K^+)$ reaction experiemnt \cite{BL_O16L},
concerning the systematic difference of $B_\Lambda^{\rm exp}$ between the emulsion and $(\pi^+, K^+)$ experiments reported in Ref. \cite{Gogami2016}.
The Fermi momentum $k_F$ [fm$^{-1}$] used in this calculation and nuclear root-mean-square radius $r_{\rm rms}$ [fm] are also shown.
In comparison, $E$, $E_x$, and $r_{\rm rms}$ are also listed for the corresponding core states.
}
  \label{Tab:hyper}
  \begin{ruledtabular}
  \begin{tabular}{ccccccccccccc}
    Hypernuclei & $k_F$ & $J^\pi$ & $E$ & $E_x$ & $r_{\rm rms}$ & $B_\Lambda$ & $B^{\rm exp}_\Lambda$ & Core nuclei & $J^\pi$ & $E$ & $E_x$ & $r_{\rm rms}$ \\
    \hline
    $^7_\Lambda$Li    & 1.05 & $1/2^+$ &  -45.69 & 0.00 & 2.28 &  5.66 &  5.58 $\pm$ 0.03 \cite{BL_C13L}  & $^6$Li & $1^+$ & -40.03 & 0.00 & 2.34  \\
                      &      & $3/2^+$ &  -45.02 & 0.67 & 2.28 &  4.99 &   &  &  &  &  &  \\
                      &      & $5/2^+$ &  -43.84 & 1.85 & 2.19 &  6.02 &   &  & $3^+$ & -37.82 & 2.21 & 2.24 \\
                      &      & $7/2^+$ &  -43.35 & 2.33 & 2.19 &  5.53 &   &  &  &  &  &  \\
    \\
    $^9_\Lambda$Li    & 1.02 & $3/2^+$ &  -54.04 & 0.00 & 2.38 &  9.20 &  8.50 $\pm$ 0.12 \cite{BL_Li9L} & $^8$Li & $2^+$ & -44.84 & 0.00 & 2.43  \\
                      &      & $5/2^+$ &  -53.71 & 0.33 & 2.37 &  8.87 &   &        &       &        &      &       \\
                      &      & $5/2^+$ &  -53.00 & 1.04 & 2.31 &  9.71 &   &        & $3^+$ & -43.29 & 1.55 & 2.36  \\
                      &      & $7/2^+$ &  -52.56 & 1.47 & 2.31 &  9.27 &   &        &       &        &      &       \\
    \\
    $^9_\Lambda$Be    & 1.12 & $1/2^+$ &  -61.67 & 0.00 & 2.39 &  6.81 &  6.71 $\pm$ 0.04 \cite{BL_Be9L} & $^8$Be & $0^+$ & -54.86 & 0.00 & 2.48 \\
                      &      & $5/2^+$ &  -58.85 & 2.82 & 2.39 &  6.89 &   &  & $2^+$ & -51.91 & 2.95 & 2.49  \\
                      &      & $3/2^+$ &  -58.80 & 2.87 & 2.39 &  6.94 &   &  &  &  &  &  \\
    \\
    $^{10}_\Lambda$Be & 1.10 &   $1^-$ &  -62.76 & 0.00 & 2.31 &  8.68 &  8.55 $\pm$ 0.18 \cite{Gogami2016} & $^{9}$Be & $3/2^-$ & -54.07 & 0.00 & 2.40 \\
                      &      &   $2^-$ &  -62.60 & 0.16 & 2.31 &  8.53 &   &          &  &  &  &  \\
                      &      &   $2^-$ &  -60.16 & 2.60 & 2.30 &  8.57 &   &          & $5/2^-$ & -51.59 & 2.49 & 2.40 \\
                      &      &   $3^-$ &  -60.00 & 2.76 & 2.30 &  8.41 &   &          &  &  &  &  \\
                      &      &   $3^-$ &  -56.18 & 2.58 & 2.26 &  8.94 &   &          & $7/2^-$ & -47.24 & 6.83 & 2.37 \\
                      &      &   $4^-$ &  -56.16 & 2.60 & 2.26 &  8.92 &   &          &  &  &  &  \\
    \\
    $^{10}_\Lambda$B  & 1.09 &   $1^-$ &  -60.76 & 0.00 & 2.35 &  8.87 &  8.89 $\pm$ 0.12 \cite{BL_Be9L} & $^{9}$B & $3/2^-$ & -51.89 & 0.00 & 2.46 \\
                      &      &   $2^-$ &  -60.58 & 0.18 & 2.35 &  8.69 &                                 &         &  &  &  &  \\
                      &      &   $2^-$ &  -58.15 & 2.61 & 2.34 &  8.72 &                                 &         & $5/2^-$ & -49.43 & 2.46 & 2.46 \\
                      &      &   $3^-$ &  -58.05 & 2.71 & 2.34 &  8.62 &                                 &         &  &  &  &  \\
%    $^{10}_\Lambda$B  & 1.04 &   $1^-$ &  -67.40 & 0.00 & 2.49 &  9.78 &  8.89 $\pm$ 0.12 \cite{BL_Be9L} & $^{9}$B & $3/2^-$ & -57.62 & 0.00 & 2.54 \\
%                      &      &   $2^-$ &  -67.26 & 0.14 & 2.49 &  9.64 &   &  &  &  &  &  \\
%                      &      &   $2^-$ &  -65.49 & 1.91 & 2.46 &  9.77 &   &  & $5/2^-$ & -55.73 & 1.90 & 2.52 \\
%                      &      &   $3^-$ &  -65.37 & 2.04 & 2.47 &  9.64 &   &  &  &  &  &  \\
    \\
    $^{11}_\Lambda$B  & 1.05 & $5/2^+$ &  -82.33 & 0.00 & 2.45 & 11.18 & 10.24 $\pm$ 0.05 \cite{BL_Be9L} & $^{10}$B & $3^+$ & -71.15 & 0.00 & 2.49  \\
                      &      & $7/2^+$ &  -81.95 & 0.38 & 2.44 & 10.80 &   &  &  &  &  &  \\
                      &      & $1/2^+$ &  -80.80 & 1.53 & 2.48 & 10.24 &   &  & $1^+$ & -70.56 & 0.59 & 2.52  \\
                      &      & $3/2^+$ &  -80.77 & 1.55 & 2.46 & 10.21 &   &  &  &  &  &  \\
                      &      & $1/2^+$ &  -79.14 & 3.19 & 2.52 & 10.43 &   &  & $1^+$ & -68.71 & 2.44 & 2.54  \\
                      &      & $3/2^+$ &  -78.88 & 3.44 & 2.52 & 10.17 &   &  &  &  &  &  \\
    \\
    $^{12}_\Lambda$C  & 1.08 &   $1^-$ &  -90.22 & 0.00 & 2.45 & 12.28 & 10.76 $\pm$ 0.19 \cite{BL_Li9L} & $^{11}$C & $3/2^-$ & -77.94 & 0.00 & 2.49 \\
                      &      &   $2^-$ &  -90.09 & 0.13 & 2.45 & 12.15 &   &  &  &  &  &  \\
                      &      &   $1^-$ &  -87.15 & 3.07 & 2.52 & 11.26 &   &  & $1/2^-$ & -75.89 & 2.05 & 2.55 \\
                      &      &   $0^-$ &  -86.93 & 3.29 & 2.52 & 11.04 &   &  &  &  &  &  \\
                      &      &   $2^-$ &  -85.70 & 4.52 & 2.49 & 11.55 &   &  & $5/2^-$ & -74.15 & 3.79 & 2.52 \\
                      &      &   $3^-$ &  -85.51 & 4.71 & 2.49 & 11.36 &   &  &  &  &  &  \\
                      &      &   $2^-$ &  -82.39 & 7.83 & 2.54 & 11.02 &   &  & $3/2^-$ & -71.37 & 6.57 & 2.54 \\
                      &      &   $1^-$ &  -82.21 & 8.01 & 2.54 & 10.84 &   &  &  &  &  &  \\
    \\
    $^{13}_\Lambda$C  & 1.10 & $1/2^+$ & -105.74 & 0.00 & 2.47 & 12.29 & 11.69 $\pm$ 0.19 \cite{BL_C13L} & $^{12}$C & $0^+$ & -92.82 & 0.00 & 2.53 \\
                      &      & $5/2^+$ & -100.98 & 4.76 & 2.52 & 12.25 &   &  & $2^+$ & -88.73 & 4.09 & 2.56 \\
                      &      & $3/2^+$ & -100.91 & 4.83 & 2.52 & 12.18 &   &  &  &  &  &  \\
    \\
    $^{15}_\Lambda$N  & 1.13 & $3/2^+$ & -125.15 & 0.00 & 2.54 & 13.70 & 13.59 $\pm$ 0.15 \cite{BL_Be9L} & $^{14}$N & $1^+$ & -111.45 & 0.00 & 2.56 \\
                      &      & $1/2^+$ & -124.98 & 0.16 & 2.54 & 13.53 &   &  &  &  &  &  \\
                      &      & $3/2^+$ & -120.23 & 4.92 & 2.58 & 12.94 &   &  & $1^+$ & -107.29 & 4.16 & 2.59 \\
                      &      & $1/2^+$ & -120.00 & 5.15 & 2.59 & 12.71 &   &  &  &  &  &  \\
    \\
    $^{16}_\Lambda$O  & 1.16 &   $1^-$ & -127.56 & 0.00 & 2.57 & 13.49 &   & $^{15}$O & $1/2^-$ & -114.07 & 0.00 & 2.57 \\
                      &      &   $0^-$ & -127.38 & 0.18 & 2.57 & 13.31 & 12.96 $\pm$ 0.05 \dag \cite{BL_O16L} &  &  &  &  &  \\
                      &      &   $1^-$ & -116.93 &10.63 & 2.62 & 12.70 &   &  & $3/2^-$ & -104.23 & 9.84 & 2.62 \\
                      &      &   $2^-$ & -116.71 &10.85 & 2.62 & 12.48 &   &  &  &  &  &  \\
  \end{tabular}
  \end{ruledtabular}
\end{table*}
%%%%%%%%%%%%%%%%%%%%%%%%%i

\subsection{Energy shifts by the addition of a $\Lambda$ particle}

In this Section, we discuss the change of the excitation energies in the hypernuclei in comparison with the core nuclei.
%{ \color{red}
In general, energy shifts are caused by spin-dependent nature of the $\Lambda N$ interaction \cite{NuclPhysA.754.48c(2005),NuclPhysA.804.84(2008)}. 
For example, the $\Lambda N$ spin-orbit interaction depending on the nuclear spin plays an essential role for energy shifts in hypernuclei. 
Recently, it is also discussed that the difference of nuclear radii in core states affects 
the energy shifts of the corresponding states in the hypernuclei \cite{PRC97.024330(2018)}.
It should be noted that both of these effects are included in our results.
%}

Figure \ref{fig:spectra2.eps} shows that in the most hypernuclei the excitation energies of the excited doublets are shifted from the corresponding core states.
For example, in $^{11}_\Lambda$B, the $(1/2^+_1, 3/2^+_1)$ and $(1/2^+_2, 3/2^+_2)$ doublets are largely shifted up in the excitation spectra compared to the $1^+_1$ and $1^+_2$ states in $^{10}$B.
In a more rigorous discussion, to remove the splitting energies of the doublets,
the difference of the centroid energies between the ground and excited doublets should be compared with the excitation energies in the core nuclei.
In $^{11}_\Lambda$B, the centroid energy differences between the ground and two excited doublets, $(1/2^+_1, 3/2^+_1)$ and $(1/2^+_2, 3/2^+_2)$,
are 1.25 MeV and 3.09 MeV, corresponding to the energy shifts of 0.66 MeV and 0.65 MeV, respectively.
The large shift up is also seen in the excited doublets of $^{12}_\Lambda$C, $^{15}_\Lambda$N and $^{16}_\Lambda$O.
On the other hand, in the Li hypernuclei, it is seen that the addition of the $\Lambda$ particle decreases the excitation energies.
For example, in $^9_\Lambda$Li, the difference of the centroid energies between the ground and $(5/2^+, 7/2^+)$ doublets is 1.12 MeV, which is smaller than the excitation energy (1.57 MeV) of the $3^+$ state in $^8$Li.

%The energy shifts caused by a $\Lambda$ particle are closely related to the difference of the nuclear radii,
%which was pointed out by Ref. \cite{PRC97.024330(2018)} based on the microscopic $\alpha$ cluster model calculation
%for $p$ shell $\Lambda$ hypernuclei.
%In Ref. \cite{PRC97.024330(2018)}, 
%{ \color{red}
It is pointed out that the energy shifts caused by a $\Lambda$ particle are 
closely related to the difference of the nuclear radii \cite{PRC97.024330(2018)}. 
In Ref. \cite{PRC97.024330(2018)}, 
based on the microscopic $\alpha$ cluster model calculation for $p$ shell $\Lambda$ hypernuclei, 
%}
it is shown that the excitation energy shift is correlated with the difference of the nuclear size.
This is because the $\Lambda$ particle in $s$ orbit penetrates into nuclear interior,
which probes nuclear density through the $\Lambda N$ interaction.
Increasing the nuclear radius makes the overlap between the $\Lambda$ particle and nucleons smaller,
which results in decreases of the $\Lambda$ binding energy,
whereas the $\Lambda$ particle is deeply bound in a spatially compact state.
The difference of the $\Lambda$ binding energies depending on the size of each state
causes the energy shifts in excitation spectra.
Similar phenomena were pointed out in deformed states,
where increasing the nuclear quadrupole deformation makes the $\Lambda$ binding energy shallower
corresponding to the smaller overlap between the $\Lambda$ and nucleons in deformed states \cite{PhysRevC.89.024310(2014),PhysRevC.92.044326(2015)}.

In the present results, we see the correlation between the energy shifts and nuclear radii.
In $^{11}_\Lambda$B,$^{12}_\Lambda$C, $^{15}_\Lambda$N and $^{16}_\Lambda$O, the excited states are shifted up by the addition of a $\Lambda$ particle, indicating that the $\Lambda$ binding energy is smaller in the excited states than the ground states.
In Tab. \ref{Tab:hyper}, we see the difference in the nuclear root-mean-square radii ($r_{\rm rms}$) and in the $\Lambda$ binding energies $B_\Lambda$ between the ground and excited states. Here, the $B_\Lambda$ is defined not only for the ground state but also excited states as the energy gain from the corresponding core state.
For example, in $^{11}_\Lambda$B, the $B_\Lambda$ values are about 10.2 MeV in the excited doublets $(1/2^+_1, 3/2^+_1)$ and $(1/2^+_2, 3/2^+_2)$, whereas $B_\Lambda = 11.18$ MeV and 10.80 MeV in the ground state doublet, which causes the energy shift up of the excited doublets.
It is found that the difference of $B_\Lambda$ corresponding to that of the nuclear radii.
In the core nucleus of $^{11}_\Lambda$B, $^{10}$B, the calculated values of $r_{\rm rms}$ are 2.52 fm and 2.54 fm in the $1^+_1$ and $1^+_2$ states, respectively, whereas $ r_{\rm rms} = 2.49$ fm in the ground state $3^+$.
In the corresponding states in $^{11}_\Lambda$B, the nuclear radii are larger in the excited doublets $(1/2^+_1, 3/2^+_1)$ and $(1/2^+_2, 3/2^+_2)$ than the ground-state doublet after the shrinkage of the system with the $\Lambda$ particle.
On the other hand, in $^{7}_\Lambda$Li and $^{9}_\Lambda$Li,
where the excited doublets are shifted down in the excitation spectra,
the nuclear radii of the excited states are smaller than the ground state.
For example, in $^{7}_\Lambda$Li, the $r_{\rm rms}$ is 2.19 fm in the excited doublet $(5/2^+, 7/2^+)$,
which is smaller than those in the ground state doublet ($ r_{\rm rms} =2.28$ fm).
The trend of the energy shift down or up in $^{7}_\Lambda$Li, $^{11}_\Lambda$B, $^{12}_\Lambda$C, and $^{15}_\Lambda$N is
consistent with the results in Ref. \cite{PRC97.024330(2018)}.

\section{Summary}

We investigate the spin-spin and spin-orbit splittings of the $p$-shell $\Lambda$ hypernuclei
within the framework of HyperAMD with the $\Lambda N$ $G$-matrix interaction ESC14+MBE.
In Ref. \cite{PRC95.044308(2017)}, it is shown that the mass dependence of the $B_\Lambda$ values are
nicely reproduced by describing the nuclear core structure
within the framework of the HyperAMD with the latest version of the YNG interaction, ESC14+MBE.
At the same time, however, this preceding analysis showed the discrepancy
in energy level order of ground-state spin doublets.
This fact tells us that there is a room to correct the spin-dependent parts of ESC14+MBE.

In this paper, we make an adjustment of the $\Lambda N$ spin-spin and spin-orbit interactions derived from ESC14+MBE.
Since the central force of the ESC14+MBE has a three range Gaussian form,
we add correction terms $\Delta V^{E} \exp \left[ - (r/\beta_2)^2 \right]$
and $\Delta V^{O} \exp \left[ - (r/\beta_2)^2 \right]$
to the even and odd state interactions of the $\Lambda N$ spin-spin force, respectively.
By the analysis and comparison with the observed data in $^{4}_\Lambda$H and $^{7}_\Lambda$Li,
we determine the strength of the correction terms as $\Delta V^{E} = 1.2$ MeV and $\Delta V^{O} = 30.0$ MeV.
The spin-orbit force of the ESC14+MBE is tuned by multiplying a correction factor $\alpha_{ALS}$ to the ALS force
to control the cancellation between the SLS and ALS forces.
In $^{9}_\Lambda$Be,
it is found that the splitting energy of the $(3/2^+, 5/2^+)$ doublet decreases as the $\alpha_{ALS}$ increases,
and thus we determine $\alpha_{ALS} = 1.9$ to reproduce the experimental value of energy spacing.

Using the tuned version of the ESC14+MBE, we have applied the HyperAMD to ten $p$-shell $\Lambda$ hypernuclei,
namely $^{7}_\Lambda$Li, $^{9}_\Lambda$Li, $^{9}_\Lambda$Be, $^{10}_\Lambda$Be, $^{10}_\Lambda$B,
$^{11}_\Lambda$B, $^{12}_\Lambda$C,  $^{13}_\Lambda$C, $^{15}_\Lambda$N, and $^{16}_\Lambda$O.
The present calculation reproduces not only the ground-state spin of these hypernuclei systematically
but also the energy spacing of the ground and excited doublets satisfactorily.

In most cases, it is remarkable that the excitation energies of the excited states
in the hypernuclei are different from those of the corresponding core states,
$i.e.$ the energy shifts occur by the addition of a $\Lambda$ particle.
It is found that the excitation energies are shifted up in the excited states with larger radii
than the ground state in the hypernuclei,
whereas the spatially compact excited states are lowered in excitation energy by adding a $\Lambda$ particle.
This is mainly due to the difference of the $\Lambda$ binding energies between the ground and excited states,
which decrease as the nuclear radius increases.

In conclusion, we have obtained successful description of $p$-shell hypernuclear structures
by making additional tuning of both even- and odd-state spin-dependent components
of $G$-matrix derived from the fundamental baryon-baryon interaction ESC14+MBE.
Thus, we have proposed a new practical set of the YNG parameters of $G$-matrix interactions
to be applied further to a wide mass region.
Using this interaction together with the hypernuclear structure models including HyperAMD, 
it will be possible to predict various phenomena caused by a $\Lambda$ particle such as nuclear structure changes in hypernuclei, 
which can be compared with existing and/or future experiments quantitatively.

%\begin{acknowledgments}
%XXX
%\end{acknowledgments}

%\section{A little more on appendixes}
%\subsection{\label{app:subsec}A subsection in an appendix}

%\newpage %Just because of unusual number of tables stacked at end
%\bibliography{XXX} % Produces the bibliography via BibTeX.

\end{document}